\documentclass[aps,amssymb,10pt,showpacs,showkeys,letterpaper]{revtex4}
\usepackage[utf8]{inputenc}
\usepackage{graphicx}
\usepackage{amsmath}
\usepackage{epsfig}
\usepackage{bm}

\usepackage{enumitem}
\usepackage{dcolumn}
\usepackage{array,multirow}
\usepackage{epstopdf}
\usepackage{amsfonts}
\usepackage{amssymb}
\usepackage{tabularx}

\usepackage{color}
\usepackage{xcolor}

\newcommand{\be}{\begin{equation}}
\newcommand{\ee}{\end{equation}}
\newcommand{\beq}{\begin{eqnarray}}
\newcommand{\eeq}{\end{eqnarray}}

\begin{document}

\title{Time like geodesics in three-dimensional\\ rotating Ho\v{r}ava AdS black hole
 }
\author{P. A. Gonz\'{a}lez}
\email{pablo.gonzalez@udp.cl}
\affiliation{Facultad de
	Ingenier\'{i}a y Ciencias, Universidad Diego Portales, Avenida Ej\'{e}rcito
	Libertador 441, Casilla 298-V, Santiago, Chile.}
\author{ Marco Olivares }
\email{marco.olivaresr@mail.udp.cl}
\affiliation{ Facultad de Ingenier\'ia y Ciencias, Universidad Diego Portales,
	Avenida Ej\'ercito Libertador 441, Casilla 298-V, Santiago, Chile.}
\author{Eleftherios Papantonopoulos}
\email{lpapa@central.ntua.gr}
\affiliation{Department of
	Physics, National Technical University of Athens, Zografou Campus
	GR 157 73, Athens, Greece.}
\author{Yerko V\'{a}squez}
\email{yvasquez@userena.cl}
\affiliation{Departamento de F\'{\i}sica, Facultad de Ciencias, Universidad de La Serena,\\
	Avenida Cisternas 1200, La Serena, Chile.}

\date{\today}

\begin{abstract}

We study the motion of particles in the background of a three-dimensional rotating Ho\v{r}ava AdS black hole that corresponds to a Lorentz-violating version of the BTZ black hole and we analyze the effect of the breaking of Lorentz invariance in such motion by solving analytically the  geodesic equations. Mainly, we find that Lorentz-violating version of the BTZ black hole posses a more rich geodesic structure, where the planetary and circular orbits are allowed, which does not occurs in the BTZ background.

\end{abstract}

\maketitle

\tableofcontents

\newpage

\section{Introduction}

The three-dimensional models of gravity are of interest because it is possible to investigate efficiently some of their properties which are shared by their higher dimensional analogs, and also exhibit interesting solutions such as particle-like solutions and black holes. In this context, three-dimensional general relativity (GR), which has no local gravitational degrees of freedom, and is Lorentz invariant presents the well know Bañados–Teitelboim–Zanelli (BTZ) black hole solution with a negative cosmological constant \cite{Banados:1992wn}. Also, it presents interesting properties at both classical and quantum levels and the BTZ solution shares several features of the Kerr black hole \cite{Carlip:1995qv}. An important issue in gravitational physics is to know the kind of orbits that test particles follow outside the event horizon of a black hole. This information can be provided by studying the geodesics around these black holes, in this context for the BTZ background it was shown that while massive particles always fall into the event horizon and no stable orbits are possible \cite{Farina:1993xw}, massless particles can escape or plunge to the horizon \cite{Cruz:1994ir}.\\

The three-dimensional Ho\v{r}ava gravity gravity \cite{Sotiriou:2011dr} admits a Lorentz-violating version of the BTZ black hole, i.e. a black hole solution with AdS asymptotics, only in the sector of the theory in which the scalar degree of freedom propagates infinitely fast \cite{Sotiriou:2014gna}. Remarkably, in contrast to GR, the three-dimensional Ho\v{r}ava gravity also admits black holes with positive and vanishing cosmological constant. Nowadays, one could think that Lorentz invariance may not be fundamental or exact, but is merely an emergent symmetry on sufficiently large distances or low energies. It has been suggested in Ref. \cite{Horava:2009uw} that giving up Lorentz invariance by introducing a preferred foliation and terms that contain higher-order spatial derivatives can lead to significantly improved UV behavior, the corresponding gravity theory is dubbed Ho\v{r}ava gravity.
It was shown that the propagation of massive scalar fields is stable in the background of rotating three-dimensional Ho\v{r}ava AdS black holes and by employing the holographic principle the different relaxation times of the perturbed system to reach thermal equilibrium were found for the various branches of solutions \cite{Becar:2019hwk}. Also, concerning to the collision of particles, it was shown that the particles can collide on the inner horizon with arbitrarily high CM energy if one of the particles has a critical angular momentum being possible the BSW process, for the non-extremal rotating Ho\v{r}ava AdS black hole. Also, while that for the extremal BTZ black hole the particles with critical angular momentum only can exist on the degenerate horizon, for the Lorentz-violating version of the BTZ black hole the particle with critical angular momentum can exist in a region from the degenerate horizon \cite{Becar:2020hix}. Concerning to the null geodesics, it was shown that for the motion of photons new kinds of orbits are allowed, such as unstable circular orbits and trajectories of first kind. Also, it was shown that an external observer will see that photons arrive at spatial infinity in a finite coordinate time \cite{Gonzalez:2019xfr}. \\

In this work we study 
 the motion of particles
in the background of a three-dimensional rotating Ho\v{r}ava AdS black hole \cite{Sotiriou:2014gna},  with the aim of analyzing the effect of  breaking the Lorentz symmetry by calculating the time like geodesic structure. We will show that Lorentz-violating version of the BTZ black hole posses a more rich geodesic structure, where the planetary orbits are allowed, which does not occur in the BTZ background.
Also, we will have
a complete knowledge of the geodesic structure for the rotating three-dimensional Ho\v{r}ava AdS black hole allowing us to understand in depth  the  Lorentz-violating effects on the  BTZ black hole. For other studies about geodesics in three-dimensional spacetimes, see \cite{Fernando:2003gg, Cruz:2013ufa, Panotopoulos:2020zvi,Kazempour:2017gho}. \\

The work is organized as follows. In Sec. \ref{Background} we give a brief review of three-dimensional rotating Ho\v{r}ava AdS black hole. In Sec. \ref{ME} we find the motion equations for particles and we present the time like geodesic structure in Sec. \ref{geodesic}. Finally, our conclusions are in Sec. \ref{conclusion}.

\section{Three-dimensional rotating Ho\v{r}ava black holes}
\label{Background}

The three-dimensional Ho\v{r}ava gravity is described in a preferred foliation by the action  \cite{Sotiriou:2011dr}
\begin{equation}
S_{H}=\frac{1}{16\pi G_{H}}\int dT d^2x N\sqrt{g}\left[L_{2}+L_{4}\right] \,,
\end{equation}
being the line element in the preferred foliation
\begin{equation}
    ds^2=N^2dT^2-g_{ij}(dx^i+N^idT)(dx^j+N^jdT)\,,
\end{equation}
where $g_{ij}$ is the induced metric  on the constant-$T$ hypersurfaces. $G_{H}$ is a coupling constant with dimensions of a length squared, $g$ is the determinant of $g_{ij}$ and the Lagrangian $L_{2}$ has the following form
\begin{equation}
 L_{2}=K_{ij} K^{ij}-\lambda K^2+\xi\left(^{(2)}R-2\Lambda\right)+\eta a_{i} a^{i}\,,
\end{equation}
where $K_{ij}$, $K$, and $^{(2)}R$ correspond to extrinsic, mean, and scalar curvature, respectively, and $a_{i}$ is a parameter related to the lapse function $N$ via $a_{i}=-\partial_{i}\ln{N}$. $L_4$ corresponds to the set of all the terms with four spatial derivatives that are invariant under diffeomorphisms. For $\lambda=\xi=1$ and $\eta=0$,  the action reduces to that of general relativity. In the infrared limit of the theory the higher order terms $L_{4}$ (UV regime) can be neglected, and the theory is equivalent to a restricted version of the Einstein-aether theory, the equivalence can be showed by restricting the aether to be hypersurface-orthogonal and the following relation is obtained
\begin{equation}
    u_\alpha=\frac{\partial_\alpha T}{\sqrt{g^{\mu\nu}\partial_\mu T \partial \nu T}}\,,
\end{equation}
where $u_\alpha$ is a unit-norm vector field called the aether, see Ref. \cite{Jacobson:2010mx} for details. Another important characteristic of this theory is that only in the sector $\eta=0$, Ho\v{r}ava gravity admits asymptotically AdS solutions \cite{Sotiriou:2014gna}. Therefore, assuming stationary and circular symmetry the most general metric is given by
\begin{equation}\label{metric2}
ds^{2}=Z(r)^2dt^{2}-\frac{1}{F(r)^2}dr^{2}-r^{2}(d\phi+\Omega(r)dt)^2~,
\end{equation}
and by assuming the aether to be hypersurface-orthogonal, it results
\begin{equation}
    u_t=\pm \sqrt{Z(r)^2(1+F^2(r)U^2(r))}\,,u_r=U(r)\,.
\end{equation}
The theory admits the BTZ ``analogue'' to the three-dimensional rotating Ho\v{r}ava black holes described by the solution
\begin{equation}
F(r)^2= Z(r)^2 =-M +\frac{\bar{J}^2}{4r^2}-\bar{\Lambda}r^2~,~
\Omega(r)=-\frac{J}{2 r^2} \,,~
    U(r) =\frac{1}{F(r)}\left(\frac{a}{r}+br\right)\,,
\end{equation}
where
\begin{equation}
\bar{J}^2=\frac{J^2+4a^2(1-\xi)}{\xi}~,~\bar{\Lambda}=\Lambda-\frac{b^2(2\lambda -\xi-1)}{\xi}~.
\end{equation}
The sign of the effective cosmological constant $\bar{\Lambda}$ determines the asymptotic behavior (flat, dS, or AdS) of the metric. Also, $\bar{J}^2$ can be negative, this occurs when either $\xi < 0$ or $\xi > 1$, $a^2 > J^2/(4(\xi-1))$.
The aether configuration for this metric is given explicitly by
\begin{equation}
u_t = \sqrt{F^2+\left(\frac{a}{r}+br\right)^2}\,,~
u_r = \frac{1}{F^2}\left(\frac{a}{r}+br\right)\,,~
u_{\phi} = 0 \,,
\end{equation}
where $a$ and $b$ are constants that can be regarded as measures of aether misalignment, with $b$ as a measure of asymptotically misalignment, such that for $b\neq 0$ the aether does not align with the timelike Killing vector asymptotically. Note that for $\xi=1$ and $\lambda=1$, the solution becomes the BTZ black hole, and for $\xi=1$ and $\lambda \neq 1$, the solution becomes the BTZ black hole with a shifted cosmological constant $\bar{\Lambda}=\Lambda- 2b^2(\lambda-1)$. However, there is still a preferred direction represented by the aether vector
field which breaks Lorentz invariance for $\lambda \neq 1$ and $b\neq 0$. The locations of the inner and outer horizons $r = r_\pm$, are given by
\begin{equation}\label{horizon}\
r_{\pm}^2=-\frac{M}{2\bar{\Lambda}}\left(1\pm \sqrt{1+\frac{\bar{J}^2\bar{\Lambda}}{M^2}} \right)~.
\end{equation}
Considering $M>0$, a negative cosmological constant $\bar{\Lambda}<0$, and $\bar{J}^2>0$ the condition $-\bar{J}^2 \bar{\Lambda}\leq M^2$ must be fulfilled for the solution represents a black hole. For $0<-\bar{J}^2\bar{\Lambda} < M^2$ the black holes  have inner and outer horizons $r_-$ and $r_+$, the extremal case corresponds to $-\bar{J}^2 \bar{\Lambda}= M^2$, while that for $\bar{J}^2< 0$ the black holes have outer horizon $r_+$, but no inner horizon $r_-$.\\

Besides the existence of inner and outer horizons,
also there are
universal horizons, which are given by \cite{Sotiriou:2014gna}
\begin{equation}\label{universal}\
 (r_{u}^\pm)^2 =\frac{M-2ab}{2(b^2-\bar{\Lambda})}\pm\frac{1}{2(b^2-\bar\Lambda)}\left[(M-2ab)^2-(4a^2+\bar{J}^2) (b^2-\bar{\Lambda})\right]^{\frac{1}{2} }\,.
\end{equation}
On the other hand, the existence of a well-defined spacelike foliation is essential in Hořava gravity. As, it was shown in Ref. \cite{Sotiriou:2014gna}, this can be achieved by imposing the condition $F^2 + (a/r + br)^2 > 0$ or
\begin{equation}
    \frac{1}{r^2}\left((b^2-\bar{\Lambda})r^4+(2ab-M)r^2+\left(\frac{\bar{J}}{4}+a^2\right)\right)>0.
\end{equation}

\begin{figure}[!h]
\begin{center}
\includegraphics[width=110mm]{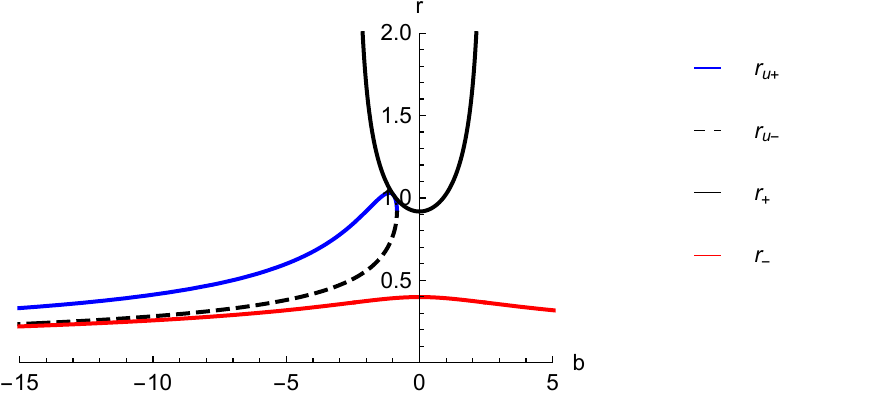}
\end{center}
\caption{The behavior of the horizons as a function of the parameter $b$,  with $M=1$, $\xi=1.2$, $\lambda=1$, $a=1$, $\Lambda=-1$, and $J=1.2$. For $b\approx -0.84$, $r_u^+=r_u^-$ \cite{Becar:2020hix}.}
\label{Hb}
\end{figure}

\begin{figure}[!h]
\begin{center}
\includegraphics[width=110mm]{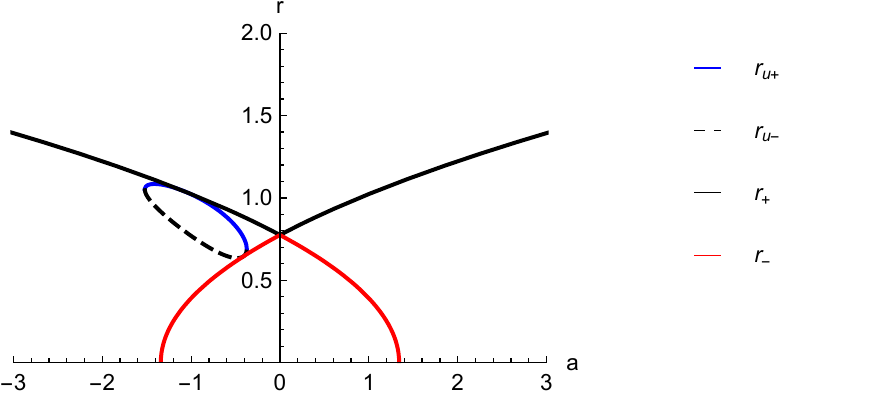}
\end{center}
\caption{The behavior of the horizons as a function of the parameter  $a$,  with $M=1$, $\xi=1.2$, $\lambda=1$, $b=1$, $\Lambda=-1$, and $J=1.2$. For $a\approx -1.52$, and $-0.37$, $r_u^+=r_u^-$ \cite{Becar:2020hix}.}
\label{Ha}
\end{figure}

The Fig. (\ref{Hb}), show the behavior of the horizons as a function of the parameter $b$, and as a function of $a$ in Fig. (\ref{Ha}),  for a choice of parameters. There are different zones, one of them is limited by $r_-$ and $r_+$; and it is described by the existence of the aether, where the roots $r_u^{\pm}$ are imaginary and therefore there are no universal horizons. Other zones are characterized by two real and distinct universal horizons inside the region between $r_-$ and $r_+$, outside $r_-$, and inside $r_+$; and an especial zone where both universal horizons coincide and is given by
\begin{equation}
    r_u^2=\frac{M-2a_{\pm}(M,\bar{J},b)b}{2(b^2-\bar{\Lambda}(b))}\,,
\end{equation}
where $a_{\pm}$ are the roots of
\begin{equation}
    \frac{(4a^2+\bar{J}^2)(b^2-\bar{\Lambda}(b))}{\xi(M-2ab)^2}=1\,.
\end{equation}
In the region between $r_u^-$ and $r_u^+$, the aether turns imaginary and the foliation cannot be extended until the singularity.

\begin{figure}[!h]
\begin{center}
\includegraphics[width=60mm]{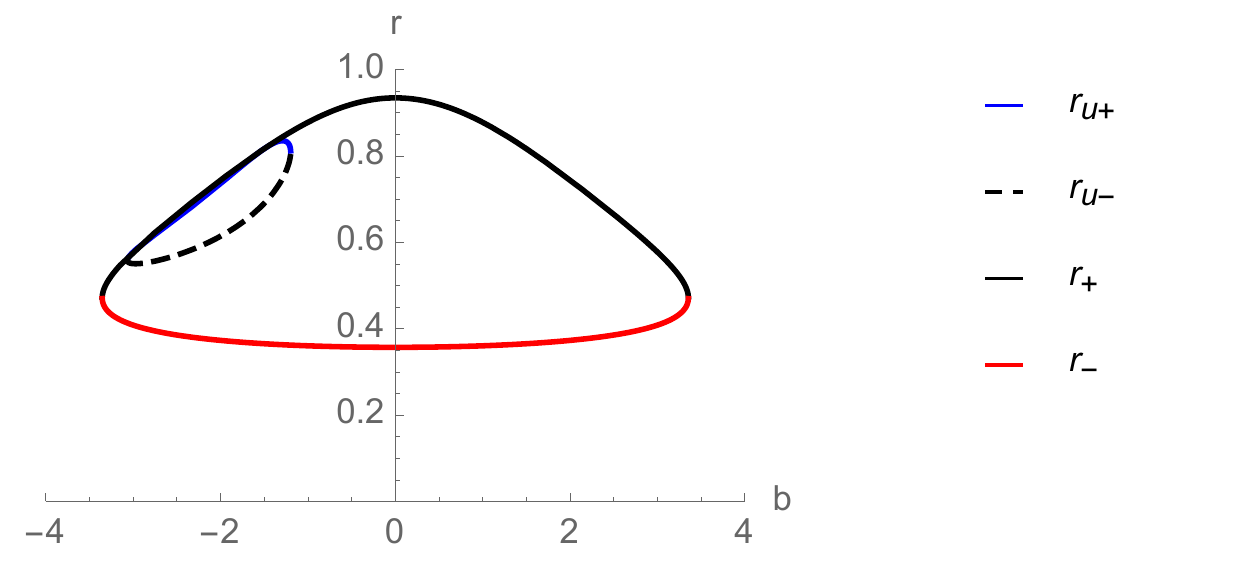}
\includegraphics[width=60mm]{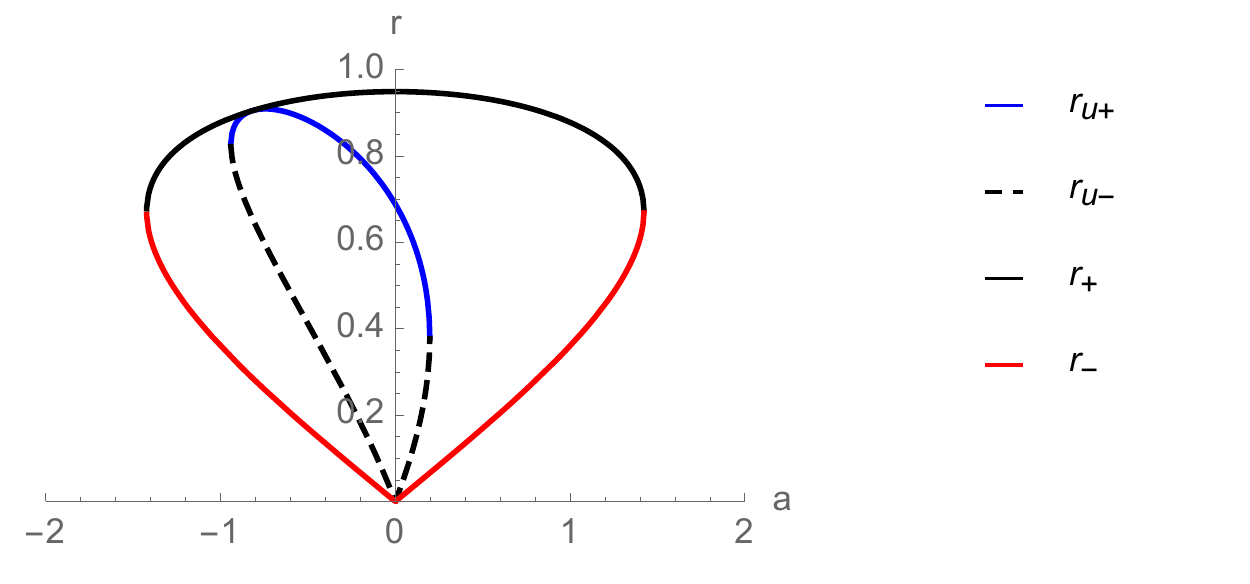}
\end{center}
\caption{The behavior of the horizons for non-rotating black holes as a function of the parameter $b$ (left panel) with $a=1$, and as a function of the parameter  $a$ (right panel) with $b=1$. Here, $M=1$, $\xi=0.9$, $\lambda=1$, and $\Lambda=-1$. 
}
\label{J0a}
\end{figure}

In order to analysis the roots of the lapse function the case $J=0$, we will consider a set of values for the parameters that satisfied the existence of two horizons $r_-$ and $r_+$, so the condition $0<-\bar{J}^2 \bar{\Lambda} < M^2$ must be satisfied, 
 and also there are not universal horizons, thereby the roots of $r_u^{\pm}$ are imaginary, as in our previous analysis. So, in order to satisfy the above condition the parameter $\xi$ must satisfied $\xi_e<\xi<\xi_c$, 
where $\xi_e$ corresponds to the  value of $\xi$ for which the black hole is extremal
\begin{equation}
\xi_{e}=\frac{2 a^2 \left(\sqrt{a^2 \left(\Lambda -2 b^2 (\lambda -1)\right)^2+b^2 (2 \lambda -1) M^2}-2 b^2 \lambda -\Lambda \right)}{M^2-4 a^2 \left(b^2+\Lambda \right)}\,,
\end{equation}
and $\xi_c$ is the value of $\xi$ for which the black hole passes from having two horizons to having one horizon and it is $\xi_c=1$ for non-rotating black holes. For $\xi>\xi_c$ the black holes are described by one horizon \cite{Becar:2019hwk}. In Fig. \ref{J0a}, we plot the behaviour of the horizons for $\xi=0.9$ for non-rotating black holes. 

In the following we will focus mainly on a set of values for the parameters that satisfied the existence of two horizons $r_-$ and $r_+$, with $a>0$, and $b>0$, where there are not universal horizons, thereby the roots of $r_u^{\pm}$ are imaginary, see Figs.  \ref{Hb}, \ref{Ha}, and \ref{J0a}. Note that under the conditions $M>0$,  $\bar{J}^2>0$, and $\bar{\Lambda}<0$, the roots of $r_u^{\pm}$ are imaginary when $M<2ab$ from Eq. (\ref{universal}).

%\newpage

\section{Equations of motion}
\label{ME}

In this section, we find the motion equations of probe massive particles around the three-dimensional Ho\v{r}ava AdS black hole. It is important to emphasize \cite{Cropp:2013sea} that in a Lorentz-violating scenario, particles will be generically coupled to the aether field generating UV modifications of the matter dispersion relations, it is more, one can also expect radiative corrections in the infrared sector but these contributions are suppressed by knows mechanisms. In our analysis we are interested on the infrared limit of the theory; so the presence of higher-order terms ($L_4$) related to the UV behaviour of the theory are ignored and in this case the theory can be formulated in a covariant fashion, and it then becomes equivalent to a restricted version of Einstein-aether theory \cite{Sotiriou:2014gna}. Since our analysis is focused in the low energy part of the theory, the interaction between the massive particle and the aether field is ignored, thus the presence of aether field only affect the background spacetime geometry. It is worth mentioning  that a similar analysis was performed in \cite{Zhu:2019ura} where the authors analyzed the evolution of the photon around the static neutral and charged aether black holes using the Hamilton-Jacobi equation. Therefore, the massive particles follow the typical geodesics in such given black holes spacetime, that can be derived from the Lagrangian of a test particle, which is given by \cite{chandra}

\begin{equation}
\mathcal{L}=\frac{1}{2}\left(g_{\mu\nu}\frac{dx^\mu}{d\tau}\frac{dx^\nu}{d\tau} \right)~.
\end{equation}
So, for the three-dimensional rotating Ho\v{r}ava AdS black hole described by the metric (\ref{metric2}), the Lagrangian associated with the motion of the test particles is given by
\begin{equation}\label{tl4}
2\mathcal{L}=-[\mathcal{F}(r)-r^2 \Omega^2(r)]\dot{t}^2+2r^2 \Omega(r)\,\dot{t}\,\dot{\phi}+
{ \dot{r}^2\over \mathcal{F}(r)}+ r^2\,\dot{\phi}^2~,
\end{equation}
where $\dot{q}=dq/d\tau$, and $\tau$ is an affine parameter along the geodesic. Here, we have defined $F(r)^2=Z(r)^2=\mathcal{F}(r)$.
Since the Lagrangian (\ref{tl4}) is
independent of the cyclic coordinates ($t,\phi$), then their
conjugate momenta ($\Pi_t, \Pi_{\phi}$) are conserved. Then, the equations of motion are obtained from
$ \dot{\Pi}_{q} - \frac{\partial \mathcal{L}}{\partial q} = 0$, and yield
\begin{equation}
\dot{\Pi}_{t} =0\,, \quad \dot{\Pi}_{r} =
-[\mathcal{F}'(r)/2-r \Omega^2(r)-r^2 \Omega'(r)]\dot{t}^{2}-
{\mathcal{F}'(r)\,\dot{r}^{2}\over 2\mathcal{F}^2(r)}
+r\,\dot{\phi}^2\,,\quad \textrm{and}\quad \dot{\Pi}_{\phi}=0~,
\label{w.11a}
\end{equation}
where $\Pi_{q} = \partial \mathcal{L}/\partial \dot{q}$
are the conjugate momenta to the coordinate $q$, and are given by
\begin{equation}
\Pi_{t} =-[\mathcal{F}(r)-r^2 \Omega^2(r)] \dot{t} +r^2 \Omega(r)\,\dot{\phi}\equiv -E~, \quad \Pi_{r}=
{ \dot{r}\over \mathcal{F}(r)}~\textrm{and}\quad \Pi_{\phi}
=r^2 \Omega(r) \dot{t} +r^2\,\dot{\phi}\equiv L~,
\label{w.11c}
\end{equation}
where $E$ and $L$ are integration constants associated to each of them.
Therefore, the Hamiltonian is given by
\begin{equation}
\mathcal{H}=\Pi_{t} \dot{t} + \Pi_{\phi}\dot{\phi}+\Pi_{r}\dot{r}
-\mathcal{L}\,.
\end{equation}
Thus,
\begin{equation}
2\mathcal{H}=-E\, \dot{t} + L\,\dot{\phi}+{ \dot{r}\over \mathcal{F}(r)}\equiv -m^2~.
\label{w.11z}
\end{equation}
where $m=1$ for timelike geodesics or $m=0$ for null geodesics.
Therefore, we obtain
\begin{eqnarray}
\label{w.14}
&&\dot{\phi}= -{1 \over (r^2-r_+^2)(r^2-r_-^2)\bar{\Lambda}}\left[ {EJ\over 2}
+L\left( {-\bar{\Lambda}r^2}
-M-{J^2-\bar{J}^2\over 4r^2}\right) \right] ~,\\
\label{w.12}
&&\dot{t}= -{ [E r^2-JL/2]\over (r^2-r_+^2)(r^2-r_-^2)\bar{\Lambda}}~,\\
\label{w.13}
&&\dot{r}^{2}=   \left( E-\frac{JL}{ 2\,r^2}\right)^2-\left( -M+\frac{\bar{J}^{2}}{4\, r^2}-\bar{\Lambda}r^2\right)  \left(m^2+\frac{L^2}{r^2}\right)=\left(E-V_-\right)\left(E-V_+\right)~,
\end{eqnarray}
where $V_{\pm}(r)$ is the effective potential and it is given by
\begin{equation}\label{tl8}
V_{\pm}(r)=\frac{JL}{ 2\,r^2}\pm\sqrt{\left( -M+\frac{\bar{J}^{2}}{4\, r^2}-\bar{\Lambda}r^2\right)  \left(m^2+\frac{L^2}{r^2}\right)}~.
\end{equation}
Since the negative branches have no classical interpretation, they are associated with antiparticles in the framework of quantum field theory \cite{Deruelle:1974zy}, We choose the positive branch of the effective potential $V = V_+$. In the next section we will perform a general analysis of the equations of motion. Note that if $\dot{t}>0$ for all $r>r_+$ the motion is forward in time outside the horizon, so from Eq. (\ref{w.12}) for $\bar{\Lambda} < 0$ the following condition must be fulfilled 
\begin{equation}
E r^2-JL/2 >0\,.
\end{equation}
On the other hand, Eq. (\ref{w.14}) can be rewritten as 
\begin{equation}
\dot{\phi} = -\frac{1}{(r^2-r_+^2)(r^2-r_-^2) \bar{\Lambda}} \left[ \frac{J}{2r^2} \left( Er^2- \frac{JL}{2}  \right)  +L F(r)^2\right]\,,
\end{equation}
due to the term $E r^2-JL/2 >0$ and $F(r)^2>0$ for $r>r_+$, when $L$, and $J$ have the same sign ($JL>0$), the term in square-brackets have not zeros outside the event horizon. However, when $L$ and $J$ have different signs ($JL<0$), the zeros can be in the relevant domain, and it would not necessarily indicate a turning point. In fact, the positive root ($R$) of the term in square-brackets in Eq. (\ref{w.14}) is
\begin{equation}
   R^2=\frac{JE }{4 \bar{\Lambda} L}-\frac{M}{2 \bar{\Lambda}} -\frac{\sqrt{(4 L M-2 JE )^2-16 \bar{\Lambda} L \left(J^2 L- \bar{J}^2 L\right)}}{8 \bar{\Lambda} L}\,,
\end{equation}
and it corresponds to the point where the angular velocity of the test particle  $\omega (r)$ 
\begin{equation}
\dot{\phi}/ \dot{t}={d\phi \over dt}={ {EJ\over 2}
+L\left( {-\bar{\Lambda}r^2}
-M-{J^2-\bar{J}^2\over 4r^2}\right) \over E r^2-JL/2}
\equiv \omega (r)\,,
\end{equation}
is null. Also, note that for a motion with $L=0$, $\omega (r)={J \over 2\,r^2}$. Thus, a particle dropped ‘straight in’ ($L=0$) from a finite distance is ‘dragged’ just by the influence of gravity so that it acquires an angular velocity ($\omega$) in the same sense as that of the source of the metric ($J$), effect called as "dragging of inertial frames".

\section{Time like  geodesics}
\label{geodesic}

\begin{figure}[!h]
	\begin{center}
		\includegraphics[width=80mm]{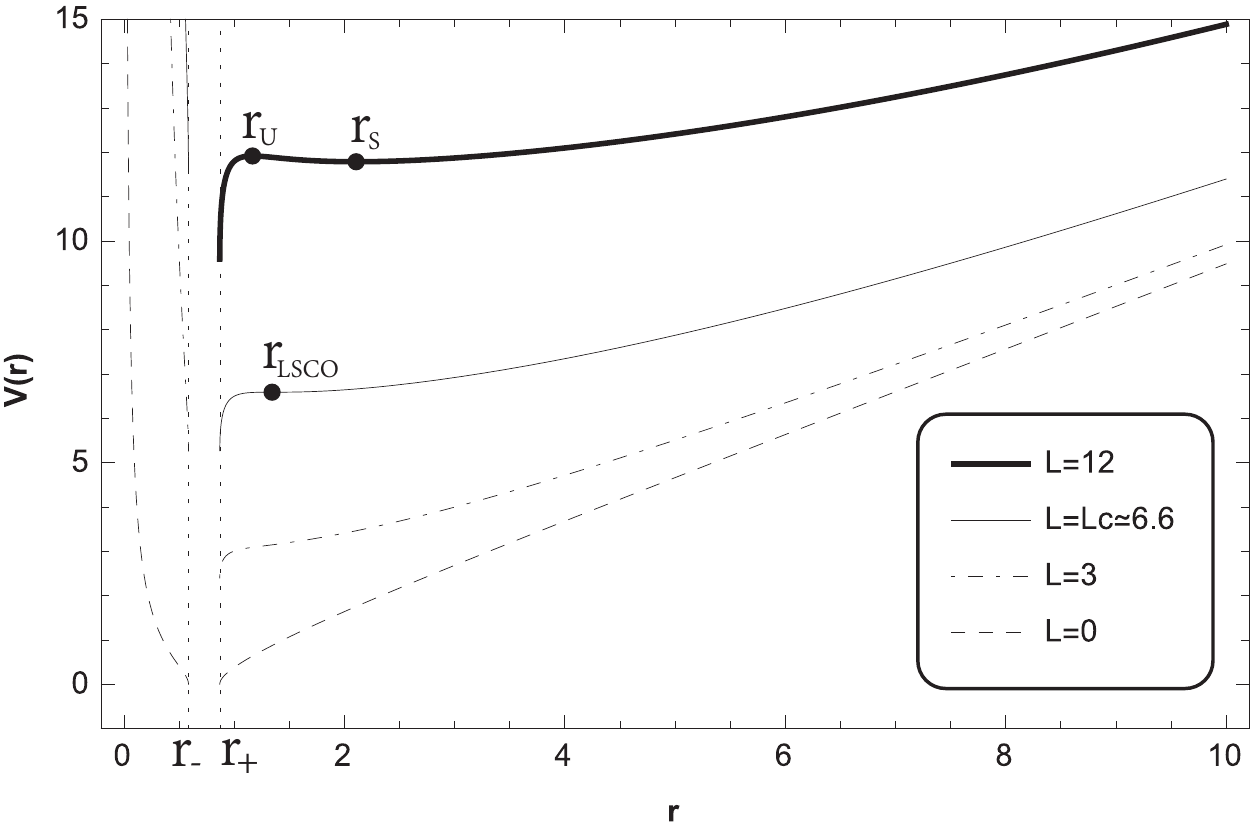}
			\includegraphics[width=80mm]{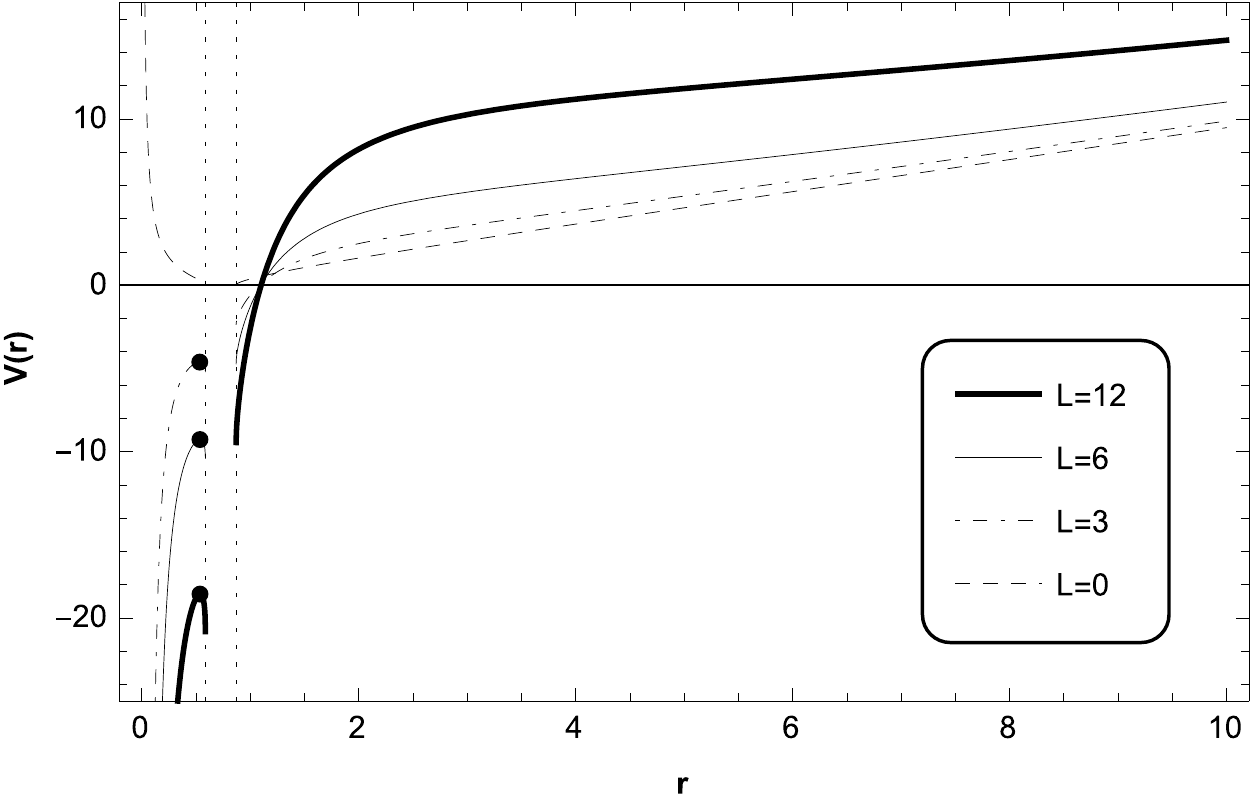}
	\end{center}
	\caption{The behavior of $V(r)$  as a function of $r$, for different values of the angular momentum of the particle $L$,
	with $M=a=b=\lambda=1$, $\xi=1.1$, $r_+\approx 0.87$
	and $\Lambda=-1$. Left panel for direct geodesics with $J=1.2$, and right panel for retrograde geodesics with $J=-1.2$. The points (right panel) indicate the extreme value of the potential that are located at $r_{ext}<r_+$.  }
	\label{f4.11}
\end{figure}

In this section we analyze the motion
of  particles, $m^2=1$, so the
effective potential is given by
\begin{equation}
V(r)=\frac{JL}{ 2\,r^2}+\sqrt{\left( -M+\frac{\bar{J}^{2}}{4\, r^2}-\bar{\Lambda}r^2\right)  \left(1+\frac{L^2}{r^2}\right)}~,
\label{t1}
\end{equation}
whose behaviour is showed in Fig. \ref{f4.11} for $J>0$, and
positive values of the angular momentum of the particle (direct geodesics), and for $J<0$, and
positive values of the angular momentum of the particle  (retrograde geodesics).  
We can observe for direct geodesics (left panel), there is a critical value of the angular momentum ($L_{LSCO}$) where the last stable circular orbit (LSCO) is present at $r=r_{LSCO}$; and for $L>L_{LSCO}$ we can distinguish four orbits, the planetary, the second kind,  the circular unstable at $r=r_U$, the circular stable at $r=r_S$, and critical orbits are allowed. We will study the existence of circular orbit in detail in subsection B. 
Note that, in the BTZ background is allowed orbits of second kind. So, the spacetime analyzed present a richer geodesic structure.  Also, for retrograde geodesics (right panel) we can observe that, for $r>r_+$, circular and planetary orbits are not allowed. Here, the trajectory always have a turning point, from which the particle plunge in the event horizon, knowing as trajectory of second kind.

The  orbit in polar coordinates is given by \footnote{In this work, we have considered a choosing of values for the parameters such that the zeros of term in square-brackets are not located in the relevant domain they are inside the event horizon.}
\begin{equation}\label{e13}
 -{r^2\over (r^2-r_+^2)(r^2-r_-^2)\bar{\Lambda}}\left[ {EJ\over 2}
+L\left( {-\bar{\Lambda}r^2}
-M-{J^2-\bar{J}^2\over 4r^2}\right) \right]
\left(\pm \frac{dr}{d\phi}\right)= \sqrt{P(r)}\,.
\end{equation}
where we have used  Eq. (\ref{w.14}) and Eq. ({\ref{w.13}}), $P(r)$ corresponds to the characteristic polynomial, and it is given by 
\begin{eqnarray}\label{tl12}
&&P(r)=r^6\,\bar{\Lambda}+r^4 \left(E^2+{L^2\bar{\Lambda}}+M\right)+r^2\left(L^2M-JEL-
\bar{J}^2/4\right)
+(J^2-\bar{J}^2)L^2/4
 \nonumber\\
&&P(r) \equiv -\bar{\Lambda}\left( -r^6+\alpha \,r^4 +\beta \, r^2
+\gamma \right) \,.
\end{eqnarray}
where
\begin{eqnarray} \label{rho0}
&&\alpha=-{E ^2+M+\bar{\Lambda}L^2 \over  \bar{\Lambda}},\\
\label{rho1}
&&\beta={JEL+\bar{J}^2/4-ML^2 \over  \bar{\Lambda}},\\
\label{rho2}
&&\gamma= - {(J^2-\bar{J}^2)L^2 \over 4 \bar{\Lambda}},
\end{eqnarray}
Therefore, we can see that depending on the nature of its roots,
we can obtain the allowed motions for this spacetime.

\newpage

\subsection{Planetary orbit}

\begin{figure}[!h]
	\begin{center}
			\includegraphics[width=80mm]{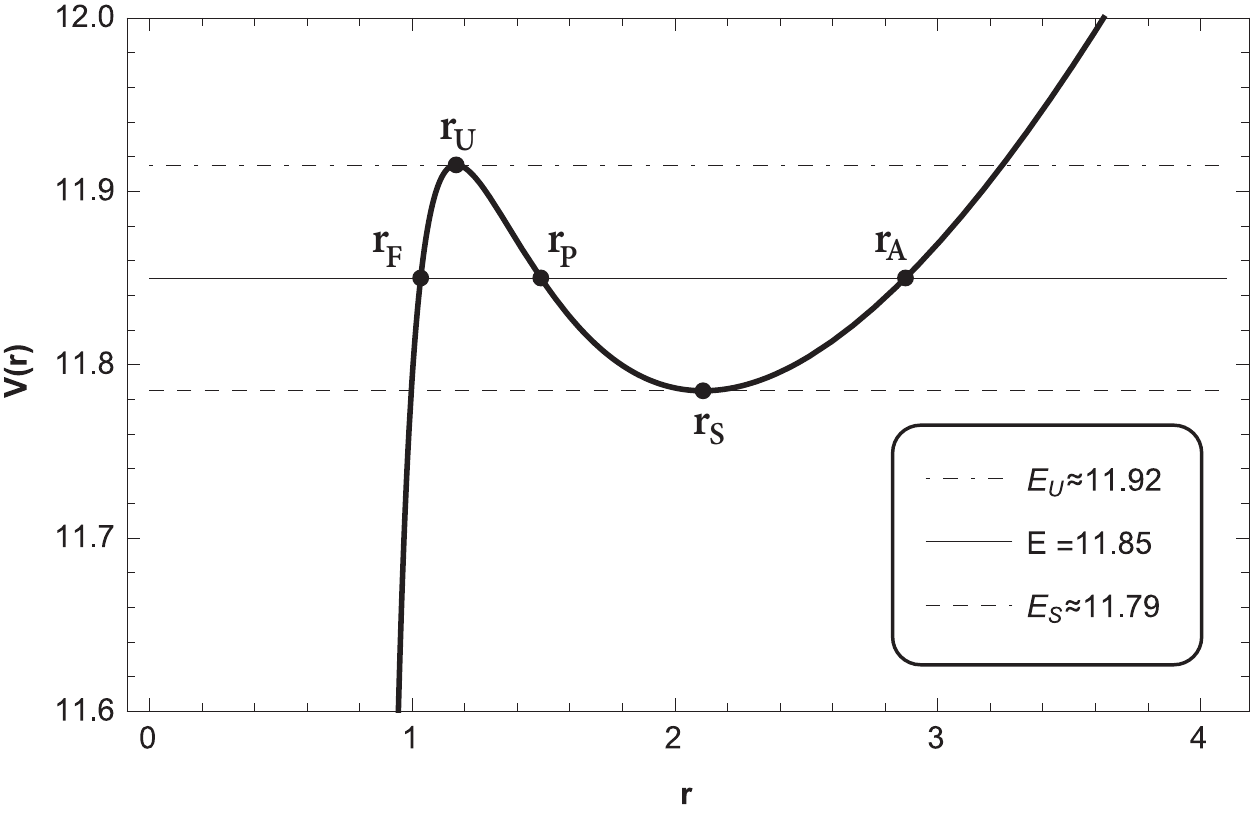}
	\end{center}
	\caption{The behavior of $V(r)$  as a function of $r$, for $L=12$, $M=a=b=\lambda=1$, $J=1.2$, $\xi=1.1$, $r_F\approx 1.03$, $r_U\approx 1.17 $, $r_P\approx 1.49 $, $r_S\approx 2.11$, $r_A\approx 2.88$, $E=11.85$, $E_S=11.79$, $E_U=11.92$, and $\Lambda=-1$.}
	\label{f4.1}
\end{figure}

The roots of  $P(r)=0$ allows to define the distance $r_P$, which corresponds
to a {\it  periastron} distance
at the trajectory,
$r_A$
which is interpreted as a
{\bf{{\it  apoastron}}} distance, and the distance $r_F$ that represent the turning point for the trajectory, see Fig. \ref{f4.1}. Thereby,
planetary orbits of the first kind occur when $L>L_C$, and the energy $E$ lies in the range $E_{S}<E<E_U$, so the radial coordinate will be $r_p<r<r_A$ for a certain value of $E$; and the planetary orbits of the second kind occurs when $E<E_U$, and $r_H<r<r_F$.

\begin{figure}[!h]
    \begin{center}
            \includegraphics[width=80mm]{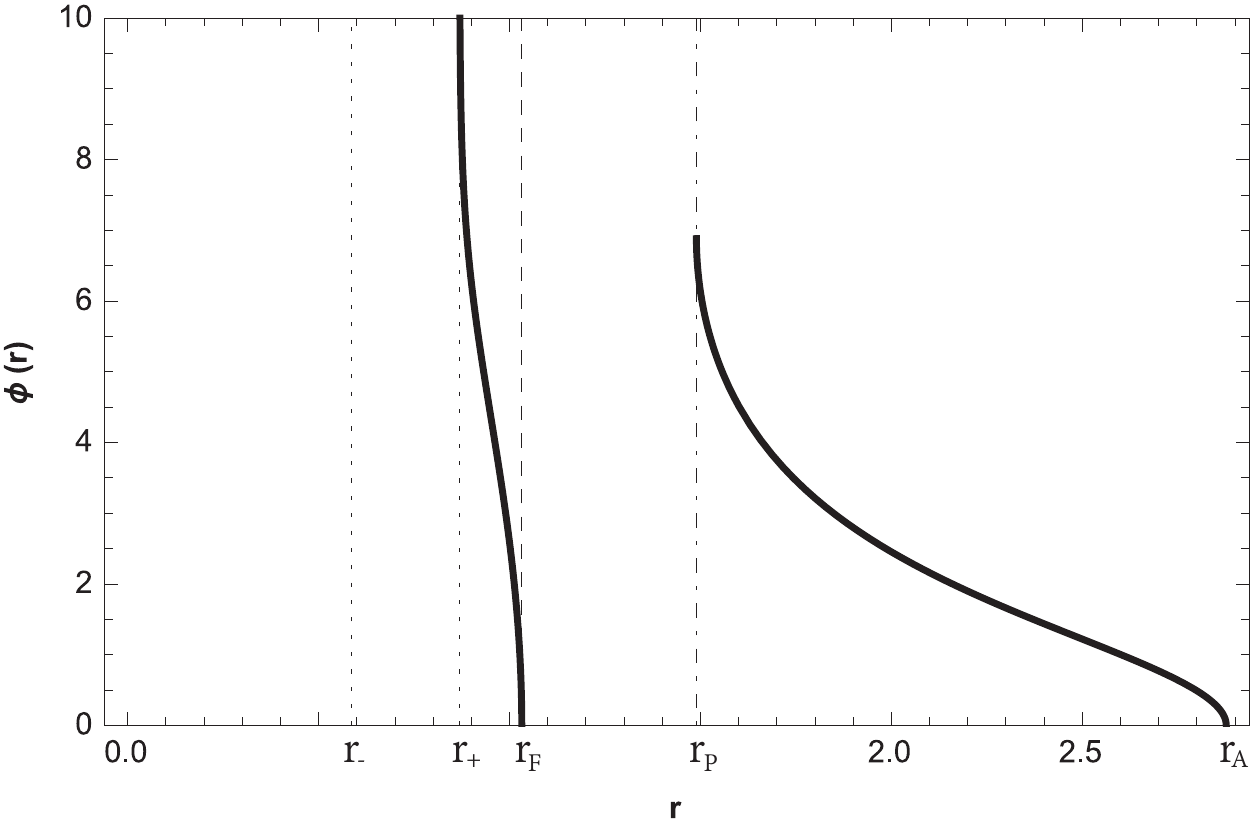}
        \includegraphics[width=60mm]{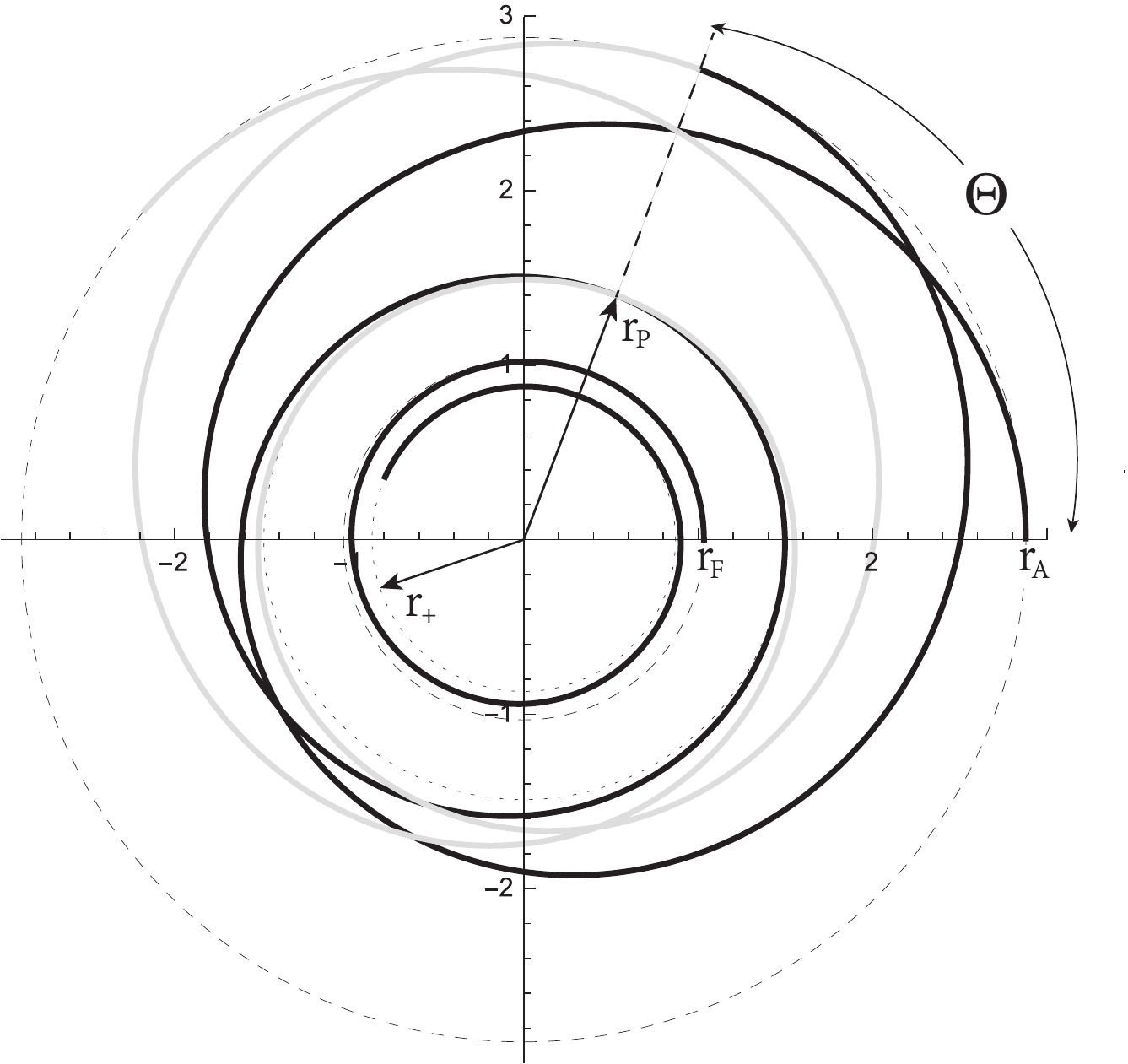}
    \end{center}
    \caption{Direct orbits. The behaviour of  $\phi(r)$ (left panel), and  $r(\phi)$ (right panel) for bounded orbits of first and second kind with $E=11.85$,  $L=12$, $M=a=b=\lambda=1$, $\Lambda=-1$, $J=1.2$, and $\xi=1.1$.}
    \label{f8}
\end{figure}

Thus, the characteristic polynomial (\ref{tl12}) can be written as
\begin{equation}\label{c10}
P(r)= -\bar{\Lambda}\left(r_{A}^2-r^2\right)(r^2-r^2_P)(r^2-r_F^2),
\end{equation}
where
\begin{equation}
r_A=\left(  {\alpha \over 3}+2\sqrt{{A \over 3}}\cos\left[ {1 \over 3}\cos^{-1}\left[ {3B \over 2}\sqrt{{3 \over A^3}}\right] \right] \right)  ^{1/2}~,
\label{rd}
\end{equation}
\begin{equation}
r_P=\left( {\alpha \over 3}+2\sqrt{{A \over 3}}\cos\left[ {1 \over 3}\cos^{-1}\left[ {3B \over 2}\sqrt{{3 \over A^3}}\right] +{4\pi \over 3} \right] \right)  ^{1/2}~
\label{rf}
\end{equation}
and
\begin{equation}
r_F=\left( {\alpha \over 3}+2\sqrt{{A \over 3}}\cos\left[ {1 \over 3}\cos^{-1}\left[ {3B \over 2}\sqrt{{3 \over A^3}}\right] +{2\pi \over 3} \right] \right)  ^{1/2}~
\label{rf}
\end{equation}
where $A={\alpha^2 \over 3}+\beta$  and $B={2\alpha^3 \over 27}+{\alpha \, \beta \over 3}+\gamma$.
Now, we will determine the angular coordinate of the trajectory for a particle that start at $r=r_A$, and it is given by the solution of the integral
\begin{equation}
\phi(r)=-\int_{r_A}^{r}  {r^2\over (-\bar{\Lambda})(r^2-r_+^2)(r^2-r_-^2)}\left[
-L\bar{\Lambda}r^2+\left( {EJ\over 2}
-ML\right) -L{J^2-\bar{J}^2\over 4r^2}\right]{d\,r\over \sqrt{P(r)}}~,
\end{equation}
 where we have used Eqs. (\ref{w.14}) and (\ref{w.13}), and whose solution is
\begin{equation}
\phi(r)=k_0\,\Psi_0(r)+K_0\left[k_+ \Psi_+(r)-k_-\,\Psi_- (r)\right]  ~,
\label{phirsolution}
\end{equation}
where
\begin{equation}
k_0={r_A\,L(J^2-\bar{J}^2)\over 8(-\bar{\Lambda})^{3/2}\sqrt{\gamma}\,r_+^2\,r_-^2}\,,
\quad
\Psi_0(r) =\wp^{-1}\left[U(r_A) \right] -\wp^{-1}\left[U(r) \right]\,,
\end{equation}
\begin{equation}
k_\pm =\frac{EJ/2-LM}{r_\pm ^2}-\bar{\Lambda}L
-\frac{L(J^2-\bar{J}^2)}{4\,r_\pm ^4}\,.
\end{equation}
This solution is plotted in Fig. \ref{f8} for direct orbits, where we can observe the trajectory of first and second kind. Note that, the coordinate ($\phi$) diverges at the event horizon.
Also, the solution (\ref{phirsolution}) allow us to determine the precession angle,
 by considering that  it  is given by
 $\Theta=2\phi_{P}-2\pi$, where $\phi_{P}$ is the angle from the apoastron to the periastron. Thus, we obtain
\begin{equation}
\Theta=2k_0\,\Psi_0(r_P)+2K_0\left[k_+ \Psi_+(r_P)-k_-\,\Psi_- (r_P)\right]  -2\pi\,.
\end{equation}
This is an exact solution for the angle of  precession, and it depends on the spacetime parameters; $M,\, J, \,\bar{J}$, and the particles motion constants, $E$ and $L$.

\begin{figure}[!h]
    \begin{center}
            \includegraphics[width=80mm]{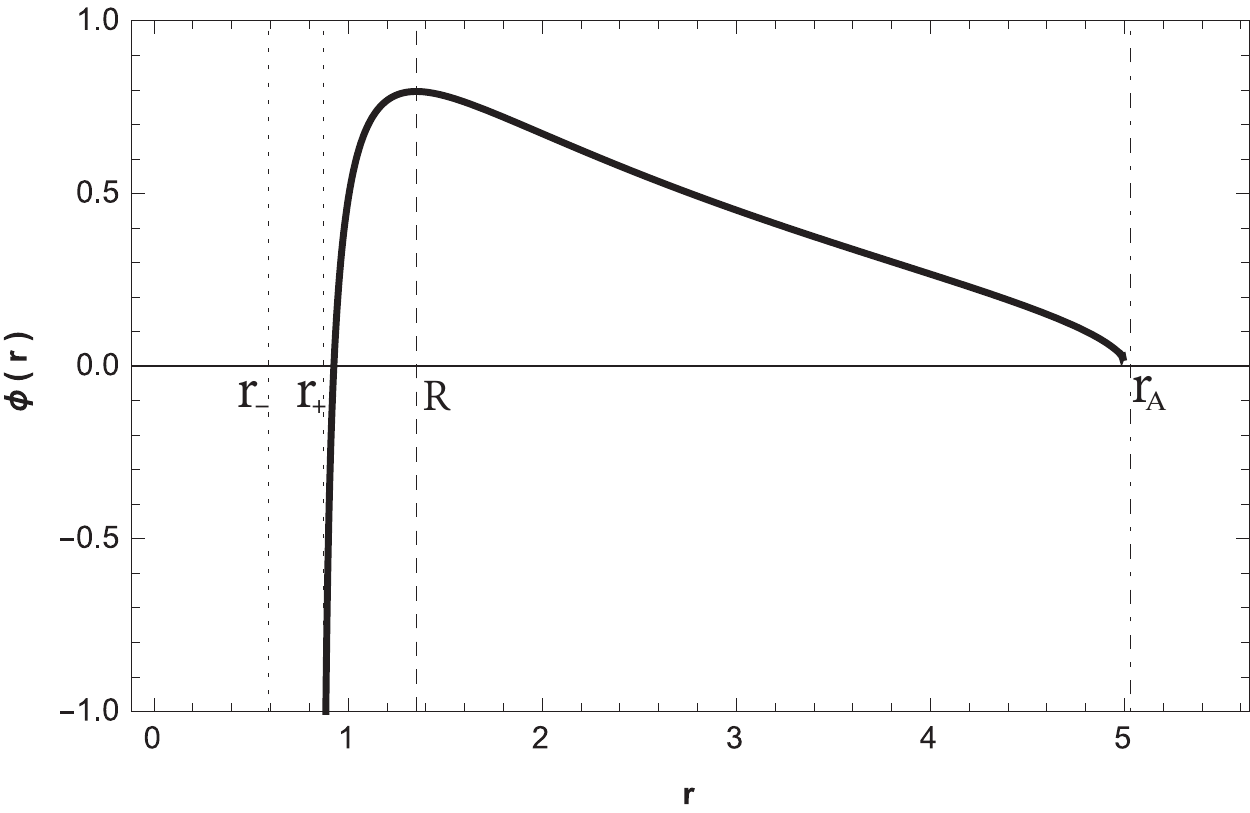}
        \includegraphics[width=80mm]{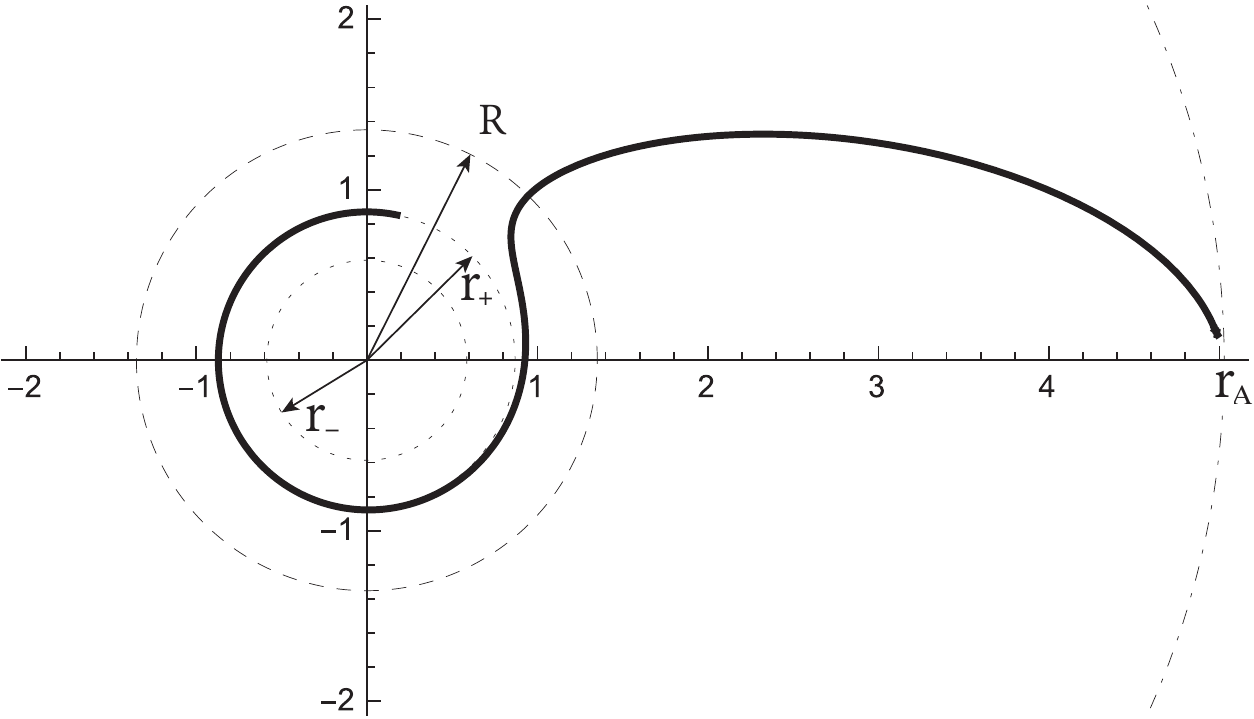}
    \end{center}
    \caption{Retrograde orbits. The behaviour of  $\phi(r)$ (left panel), and  $r(\phi)$ (right panel) for bounded orbits of second kind with $E=11.85$,  $L=12$, $M=a=b=\lambda=1$, $\Lambda=-1$, $J=-1.2$, $R=1.35$, and $\xi=1.1$.}
    \label{f6}
\end{figure}

In Fig. \ref{f6}, we plot the behaviour of the retrograde orbits, clearly they corresponds to second kind trajectories. Here, we can observe the effect of "dragging of inertial frames". The test particle start at $r=r_A$, then at $r=R$ the angular velocity is null, and $\phi(r)$ is maximum, after that the angle $\phi(r)$ decreases and tends to $-\infty$, when $r\rightarrow r_+$.

\begin{figure}[!h]
    \begin{center}
        \includegraphics[width=70mm]{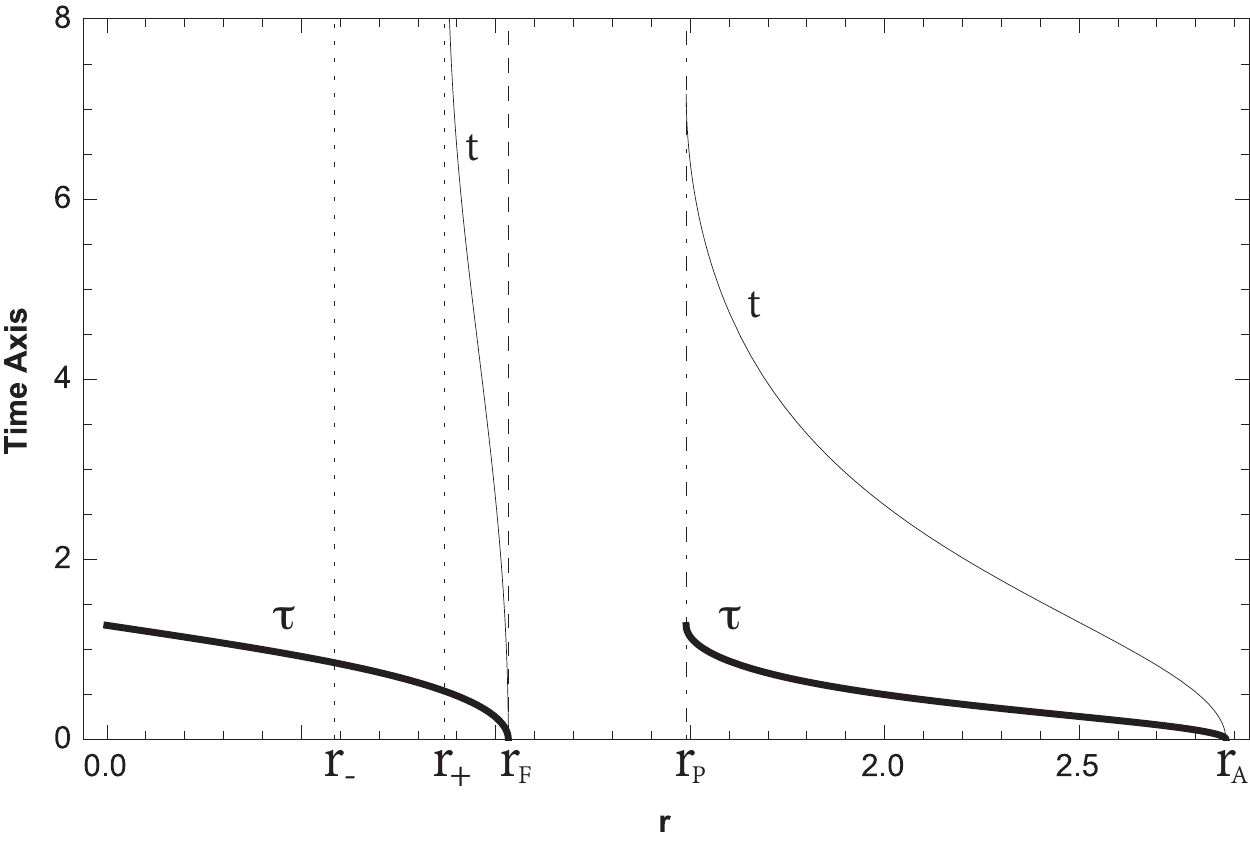}
         \includegraphics[width=70mm]{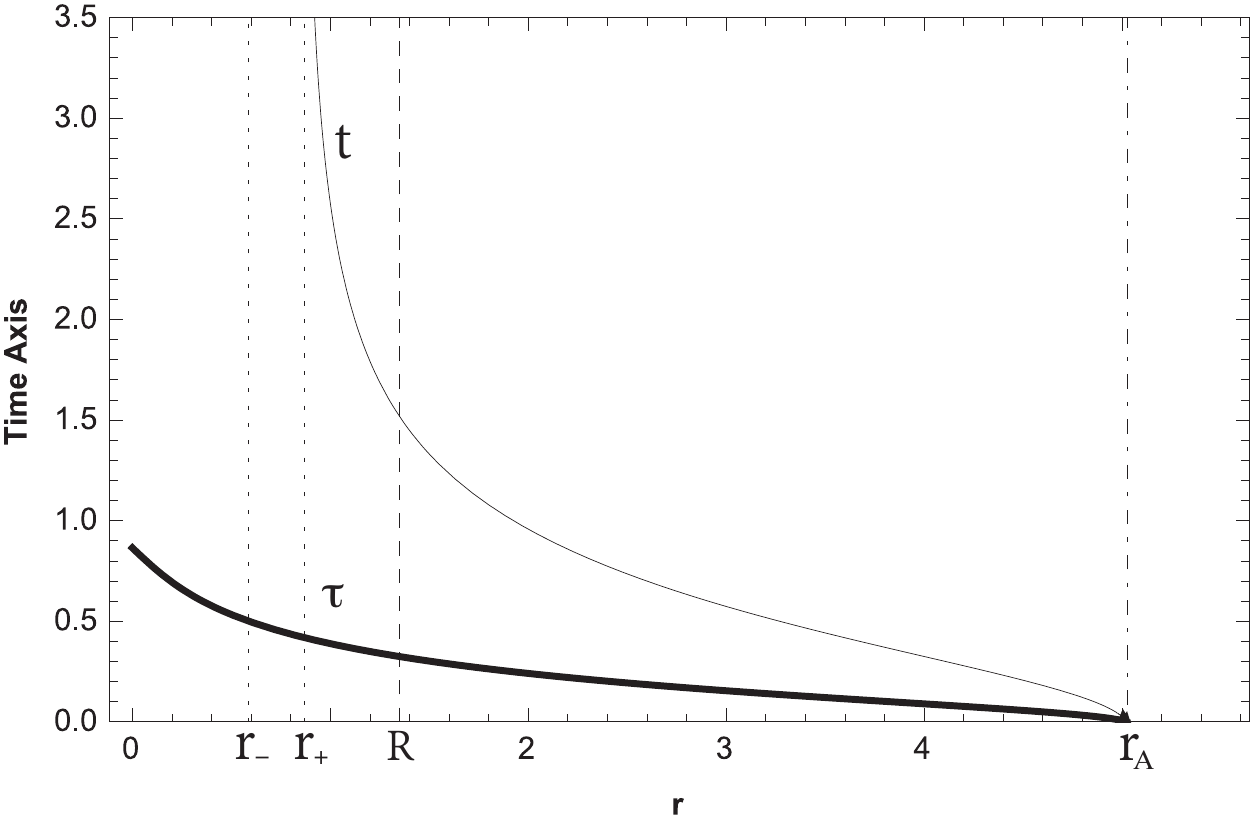}
    \end{center}
    \caption{The behaviour of the coordinate time $(t)$ and the proper time  $(\tau)$ along an bounded time-like geodesic described by a test particle, starting at $r_A$, and   $r_F=1.03$ with $r_P=1.49$, $R=1.35$, $L=12$, $M=a=b=\lambda=1$, $\Lambda=-1$, $E=11.85$, $\xi=1.1$, $r_+\approx 0.87$, and $r_{-} \approx 0.59$.
    Left panel for direct orbits with $J=1.2$,and $r_A=2.88$, and right panel for retrograde orbits  with $J=-1.2$, and $r_A=5.03$.}
    \label{tauyt1}
\end{figure}

Now, in order to determine the proper and coordinate period of rotation of the trajectories we present the solution for both times. The proper time ($\tau$), is given by the solution of the integral
\begin{equation}
\tau(r)=-\int_{r_A}^{r} {r^2\,d\,r\over \sqrt{P(r)}}~,
\end{equation}
where we have used Eq. (\ref{w.13}) and we have considered as initial conditions that the particles are at $r=r_A$ when $\phi=t=\tau=0$. Thus, we obtain % yields
%the solution
\begin{equation}
\tau(r)={r_A^3\over 8\sqrt{(-\bar{\Lambda})\,\gamma}}\left(\Psi[U(r),\Omega]-
\Psi [U(r_A),\Omega]\right)  ~,
\label{mr.3}
\end{equation}
where
\begin{equation}
\Psi(U,\Omega)=\frac{1}{\wp^{'}(\Omega)}
\left[2\zeta(\Omega)\wp^{-1}(U)
+\ln\left|\frac{\sigma[\wp^{-1}(U)-\Omega]}
{\sigma[\wp^{-1}(U)+\Omega]}\right|
\right]\,,
\end{equation}
and
\begin{equation}
U(r)={r_A^2\over 4r^2}+\frac{\beta \,r_A^2 }{12\gamma}\,,
\qquad U(r_A)={1\over 4}+\frac{\beta \,r_A^2 }{12\gamma}\,,
\end{equation}
\begin{equation}
\Omega=\wp^{-1}\left[\frac{\beta \,r_A^2 }{12\gamma}\right]
\,,
\end{equation}
\begin{equation}
g_2=\frac{r_A^4}{4}\left[\frac{\beta^2 }{3\gamma ^2}-\frac{\alpha }{\gamma }\right]\,,
\quad
g_3=\frac{r_A^6}{16}\left[\frac{\alpha \,\beta}{3\gamma ^2}-\frac{2\beta^3}{27\gamma ^3}+\frac{1}{\gamma }\right]\,.
\end{equation}
In Fig. \ref{tauyt1}, we show the behaviour of the proper time as a function of $r$ for direct orbits (left panel), we can observe, for the trajectory of first and second kind the particle arrives in a finite proper time to $r_P$, and to reach the singularity,
respectively. Also, the period of a revolution according to the proper time is $T_{\tau}=2{\tau}(r_P)$. Concerning to the retrograde orbits (right panel), for the trajectory of  second kind the particle arrives in a finite proper time to the singularity.

On the other hand, by considering Eqs. (\ref{w.12}) and  (\ref{w.13}), and as initial condition that the particles are at $r=r_A$ when $\phi=t=\tau=0$. The coordinate time ($t$) is
\begin{equation}
t(r)=-\int_{r_A}^{r}  {r^2\left[E r^2-JL/2\right]\over (-\bar{\Lambda})(r^2-r_+^2)(r^2-r_-^2)}{d\,r\over \sqrt{P(r)}}~,
\end{equation}
whose solution is
\begin{equation}
    t(r)=K_0\left[\left( E-{J\,L\over 2r_+^2}\right) \Psi_+(r)-\left( E-{J\,L\over 2r_-^2}\right)\,\Psi_- (r)\right] ~,
    \label{phi1}
\end{equation}
where $K_0={r_A^3\over 8(-\bar{\Lambda})^{3/2}\sqrt{\gamma}(r_+^2-r_-^2)}$, and
\begin{equation}
\Psi_\pm(r)=\Psi[U(r_A),\Omega_\pm]-
\Psi [U(r),\Omega_\pm]\,,
    \quad
\Omega_\pm(r)=\wp^{-1}\left[{r_A^2\over 4r_\pm^2}+\frac{\beta \,r_A^2 }{12\gamma}\right]\,.
\end{equation}
Also, in Fig. \ref{tauyt1}, we show the behaviour of the 
%proper and 
coordinate time as a function of $r$. For direct orbits, we can observe, for the trajectory of first and second kind the particle arrives in a finite %coordinate time 
%proper time 
and infinity coordinate time to $r_P$, and $r_+$, respectively. Also, the period of a revolution according to the coordinate time is $T_{t}=2{t}(r_P)$. Concerning to the retrograde orbits (right panel), for the trajectory of second kind the particle arrives in an infinite coordinate time to $r_+$. Also, we can observe that the zero located at $R$ does not affect to the proper and coordinate times.

\newpage

\subsection{Circular orbits}
The effective potential $V(r)$ has to exhibit  extrema for fixed values of radial coordinate, $r=r_{c.o.}$, when
\begin{equation}\left.  \frac{d V(r)}{dr}\right|_{r_{c.o.}}=0\,.\label{C.2}
\end{equation}
Now, for simplicity, we write the effective potential as
\begin{equation} V(r)= \frac{JL }{2\,r^2}+ \sqrt{\mathcal{F}(r)+\,  L^2\,\frac{\mathcal{F}(r)}{r^{2}}}\,. \label{C.21}\end{equation}
Therefore, using Eq. (\ref{C.21}) into Eq. (\ref{C.2}) yields
\begin{equation}\left. \left[r^3\mathcal{F}'(r)+\,  L^2\left[ r\,\mathcal{F}'(r)-2\mathcal{F}(r) \right] -2JL\sqrt{\mathcal{F}(r)+\,  L^2\,\frac{\mathcal{F}(r)}{r^{2}}}\right]\right| _{r_{c.o.}}=0\,. \label{C.22}\end{equation}
Notice that this equation leads to a polynomial of twelfth grade given by, 
\begin{eqnarray}
\label{P12}
\nonumber &&r_{c.o.}^{12}+\frac{\bar{J}^2-4L^2 M}{2\bar{\Lambda}}r_{c.o.}^8+\frac{L^2 \left(J^2+\bar{J}^2\right)}{\bar{\Lambda}}r_{c.o.}^6+\\
\nonumber &&\frac{16 J^2 L^2 \left(\bar{\Lambda} L^2+M\right)+\bar{J}^2 \left(\bar{J}^2-8 L^2 M\right)+16 L^4 M^2}{16 \bar{\Lambda}^2}r_{c.o.}^4+\\
&&\frac{L^2 \left(J^2-\bar{J}^2\right) \left(4 L^2 M-\bar{J}^2\right)}{4 \bar{\Lambda}^2}r_{c.o.}^2-\frac{\bar{J}^2 L^4 \left(J^2-\bar{J}^2\right)}{4 \bar{\Lambda}^2}=0\,,
\end{eqnarray}
so, it is possible to find the roots numerically. 
On the other hand, the condition (\ref{C.22}) allows to obtain the angular momentum for the stable $L_{c.o.}=L_S$ at $r_{c.o.}=r_S$ and for the unstable circular orbits $L_{c.o.}=L_U$ at $r_{c.o.}=r_U$, which yields
\begin{equation} \mathcal{A}\,L_{c.o.}^{4}-\mathcal{B}\,L_{c.o.}^{2}+\mathcal{C}
=0\,,\label{C.23}\end{equation}
\noindent where
\begin{eqnarray}
\mathcal{A}&=&\left. \left[4\left( M^2+\bar{\Lambda}J^2\right) +\left(J^2 -\bar{J}^2\right) \left( \frac{4M}{r^{2}}-\frac{\bar{J}^2}{r^{4}}\right) \right]\right| _{r_{c.o.}}\,,
\nonumber\\
\mathcal{B} &=&\left. \left[ \frac{\bar{J}^2}{r^{2}}\left(J^2 -\bar{J}^2\right) -2M\left(2J^2 -\bar{J}^2\right)-4\bar{\Lambda}r^2\left( J^2+\bar{J}^2-2Mr^2\right) \right]\right| _{r_{c.o.}}\,,
\nonumber\\
\mathcal{C} &=&\left. \left[ 4\bar{\Lambda}^2r^8+2\bar{\Lambda}\bar{J}^2r^4+\frac{\bar{J}^2}{4}\right]\right| _{r_{c.o.}}\,.
\end{eqnarray}
Thus, the real solution of the quartic equation for $L_{c.o.}$ is
\begin{equation} L_{c.o.}^2=  \left. \left[\frac{ \mathcal{B}-\sqrt{ \mathcal{B}^{2}-4\, \mathcal{A}\, \mathcal{C}}}{2\, \mathcal{A}}\right]\right| _{r_{c.o.}}, \label{T.25}\end{equation}
\noindent and  the energy  is given by  $E_{c.o.} = JL_{c.o.}/2\,r^2_{c.o.}+ \sqrt{\mathcal{F}(r_{c.o.})+\,  L_{c.o.}^2\,\mathcal{F}(r_{c.o.})/r_{c.o.}^2}$.

\begin{figure}[!h]
	\begin{center}
		\includegraphics[width=60mm]{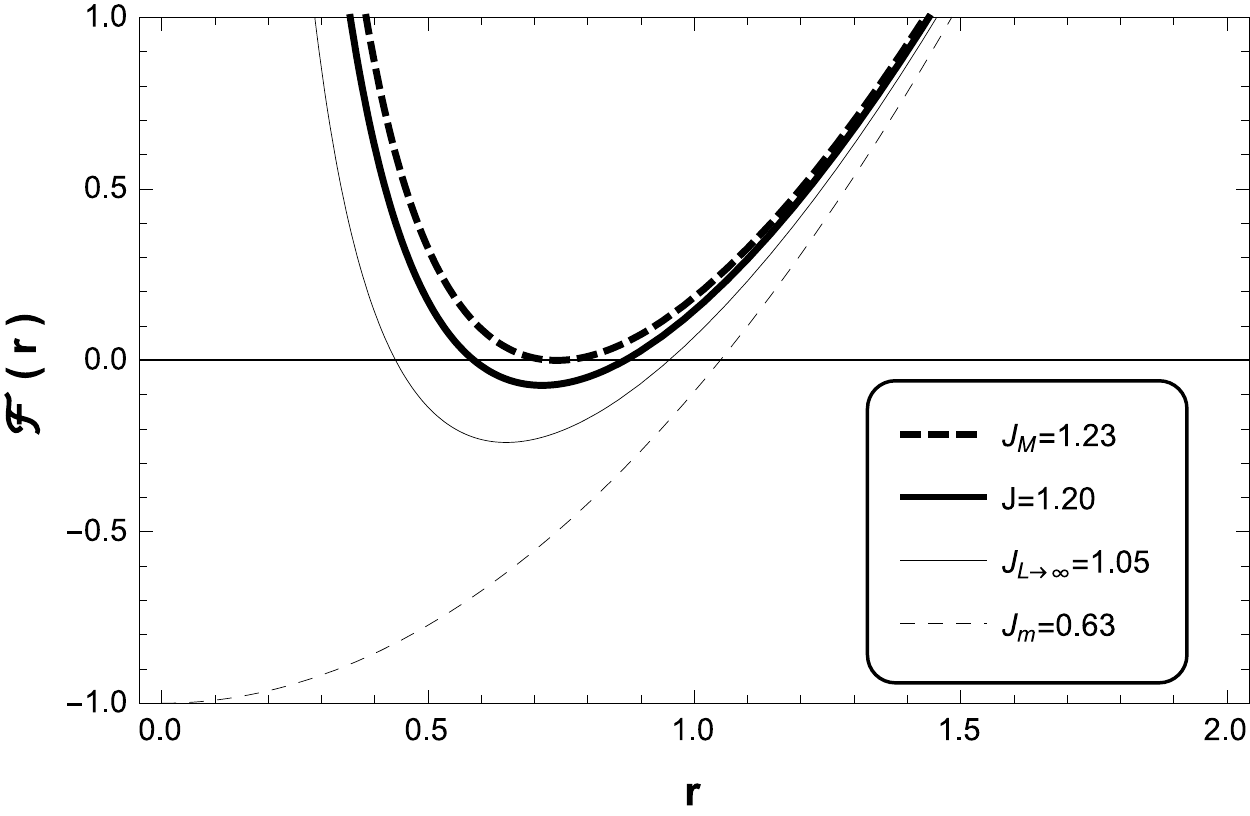}
		\includegraphics[width=60mm]{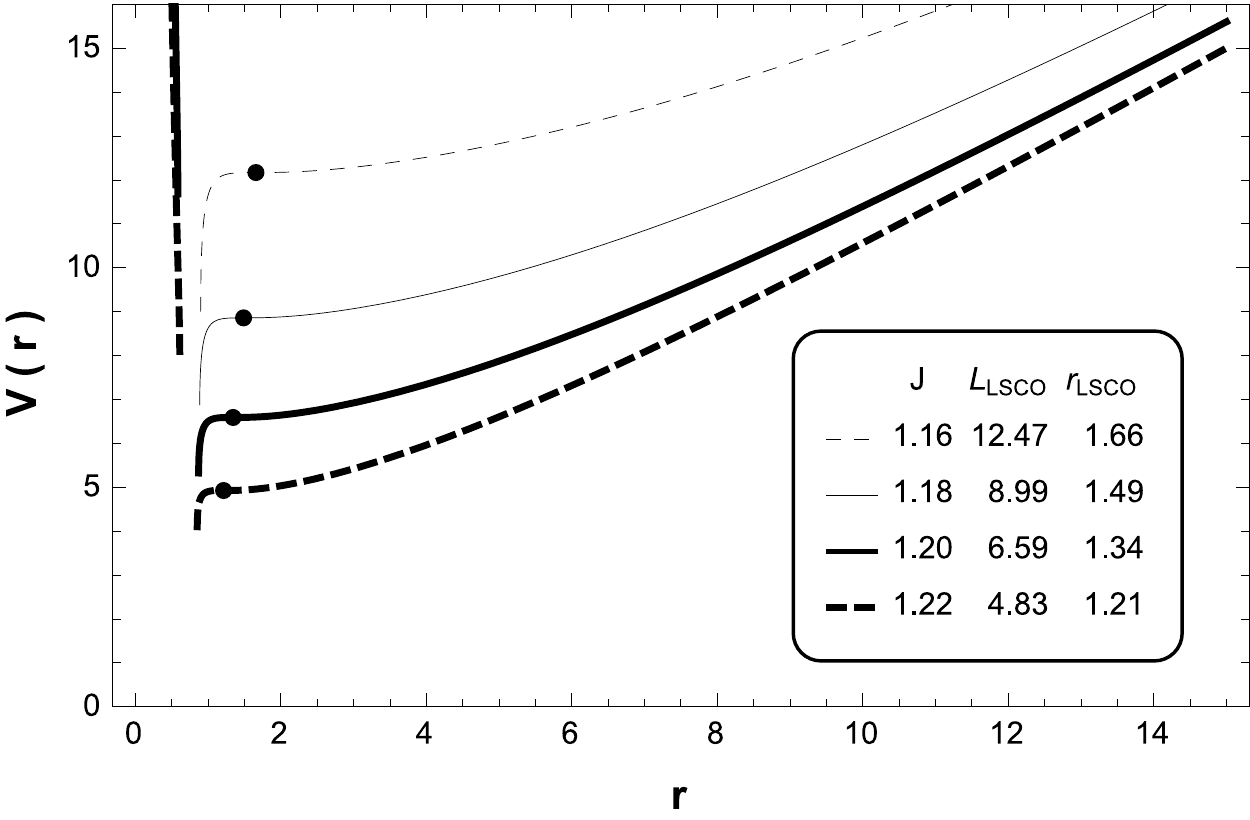}
	\end{center}
	\caption{The behaviour of the lapse function (left panel), and the behavior of $V(r)$ (right panel) as a function of $r$, for different values of the angular momentum of the black hole $J$,
	with $M=a=b=\lambda=1$, $\xi=1.1$,  and 
	$\Lambda=-1$. For $J=J_m=2a\sqrt{\xi-1}$ the black hole has one horizon at $r=r_+=1.05$, in the range $J_m<J<J_M$ the black hole has two horizons at $r=r_-=0.40$, and $r=r_+=0.97$, and for $J=J_M=\sqrt{4a^2\bar{\Lambda}(\xi-1)-M^2\xi\over \bar{\Lambda}}$ the black hole is extremal at $r=r_e=0.74$. $J=J_{L\rightarrow \infty}=\sqrt{-M^2/\bar{\Lambda}}$ corresponds to the value of $J$ for which  the minimum radius of the stable circular orbit diverges, see Eq. (\ref{RU}).}
	\label{potFV}
\end{figure}

Now, in Fig. \ref{potFV}, we show the behaviour of the lapse function and the effective potential for different values of $J$. We can observe that $J$, can not take big values due to the spacetime became in a naked singularity, and for a very small value of $J$ the black hole present only one event horizon, because the $\bar{J}^2$ became negative. 
So, 
%as the referee mentioned, 
only in spacetimes with sufficiently big
$|J|$ do the circular co-rotating orbits in the domain $r>r_+$ exist. \\

\begin{figure}[!h]
	\begin{center}
		\includegraphics[width=60mm]{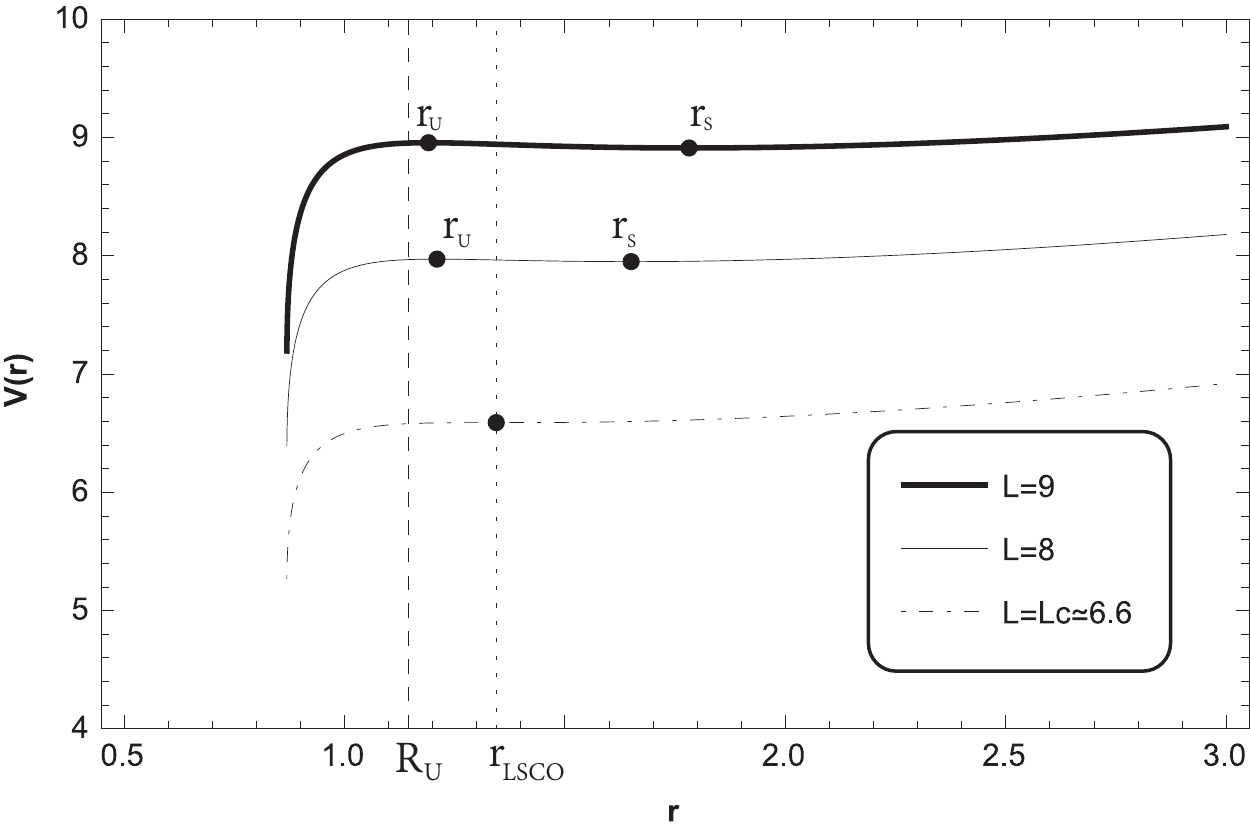}
	\end{center}
	\caption{The behavior of $V(r)$  as a function of $r$, for different values of the angular momentum of the particle $L$,
	with $M=a=b=\lambda=1$, $\xi=1.1$, $r_+\approx 0.87$, $J=1.2$,
	$\Lambda=-1$, $R_U=1.15$ and $r_{LSCO}=1.34$. For $L=8$, $r_U=1.21$, and $r_S=1.65$. For $L=9$, $r_U=1.19$, and $r_S=1.78$. Here, $L_c$=$L_{LSCO}$.}
	\label{potOC}
\end{figure}

Also, the behaviour of the effective potential for different values of angular momentum $L$, is shown in Fig. \ref{potOC}. We can observe that for $L>L_{LSCO}$, when the angular momentum increases the radius of the stable circular orbit increases, and the radius of the unstable circular orbit ($r_U$) decreases. Also, when $L\rightarrow \infty$ the $r_U \rightarrow R_U$, is given by      
\begin{equation}
\label{RU}
R_U=\left[{-M(J^2 - \bar{J}^2)- 
  |J|\sqrt{(J^2 - \bar{J}^2)(M^2 + \bar{\Lambda}\bar{J}^2)}\over  2(M^2 + \bar{\Lambda}J^2)}\right]^{1/2}\,,    
\end{equation}
that corresponds to the minimum radius for this orbit. Note that the above equation diverges at $J=J_{L\rightarrow \infty}=\sqrt{-M^2/\bar{\Lambda}}$. Also, for $J^2<-M^2/\bar{\Lambda}$, $R_U$ became imaginary. On the other hand, the radius of the last stable circular orbit must satisfied $\left. V'\right|_{r_{LSCO},L_{LSCO}}=0$ and  $\left. V'' \right|_{r_{LSCO},L_{LSCO}}=0$, and it is given by 
\begin{equation}
r_{LSCO}=\left[{L_{LSCO}^2(J^2-\bar{J}^2)\over 4(-\bar{\Lambda})}\right]^ {1/6}\,.
\end{equation}
Therefore, in the range $J_{L\rightarrow \infty}<J<J_M$ the black hole has two horizons and the effective potential allows circular and planetary orbits for $L>L_{LSCO}$.

\newpage

Also, it is possible to determine the periods of revolution of the circular orbits, both stable and unstable, with respect to the proper time $\tau$, $T_{\tau}=2\pi/\dot{\phi}(r_{c.o.})$, and coordinate time $t$, $T_{t}=T_{\tau}\,\dot{t}(r_{c.o.})$. Thereby, the period of a revolution according to the proper time is
\begin{equation}\label{ptau}
T_{\tau}= {4\pi (-\bar{\Lambda})  r_{c.o.}(r_{c.o.}^2-r_+^2)(r_{c.o.}^2-r_-^2) \over J\sqrt{(r_{c.o.}^2+L_{c.o.}^2)\mathcal{F}(r_{c.o.})}+2L_{c.o.}r_{c.o.}
\mathcal{    F}(r_{c.o.})}\,,
\end{equation}
and the period according to the coordinate time is
\begin{equation}\label{pte}
T_{t}= {4\pi  r_{c.o.}^2\sqrt{(r_{c.o.}^2+L_{c.o.}^2)\mathcal{F}(r_{c.o.})}\over J\sqrt{(r_{c.o.}^2+L_{c.o.}^2)\mathcal{F}(r_{c.o.})}+2L_{c.o.}r_{c.o.}
    \mathcal{    F}(r_{c.o.})}\,.
\end{equation}

On the other hand, Taylor expanding the effective potential around $r=r_S$, one can write $
V(r)=V(r_s)+V'(r_S)(r-r_S)+{1\over2}V''(r_S)(r-r_S)^2+...,
$
where$\,'$ means derivative with respect to the radial coordinate.
Obviously, in these orbits $V'(r_S)=0$, so, by defining the {\it smaller}
coordinate $x=r-r_S$, together with {\it the epicycle frequency}
$\kappa^2=V''(r_S)$ \cite{RamosCaro:2011wx},  we can rewrite the above equation as $ V(x)\approx E_S+\kappa^2\,x^2/2$
where  $E_S$  is the energy of the particle in the stable circular orbit. Also, it is easy to see that test particles satisfy the harmonic equation of motion $ \ddot{x}= -\kappa ^2 x$. Therefore, the epicycle frequency is given by
\begin{equation}\label{kappa}
\kappa^2=\left.{JL_{c.o.}\over r^4}\left[ 3+{r^4\mathcal{F}''+r^2L_{c.o.}^2\mathcal{F}''-4r\mathcal{F}'L_{c.o.}^2+6L_{c.o.}^2\mathcal{F}\over r^3\mathcal{F}'+r\mathcal{F}'L_{c.o.}^2-2L_{c.o.}^2\mathcal{F}}-{r^3\mathcal{F}'+r\mathcal{F}'L_{c.o.}^2-2L_{c.o.}^2\mathcal{F}\over2 (r^2\mathcal{F}+L_{c.o.}^2\mathcal{F})}\right] \right| _{r_{S}}\,.
\end{equation}

\begin{figure}[!h]
	\begin{center}
		\includegraphics[width=60mm]{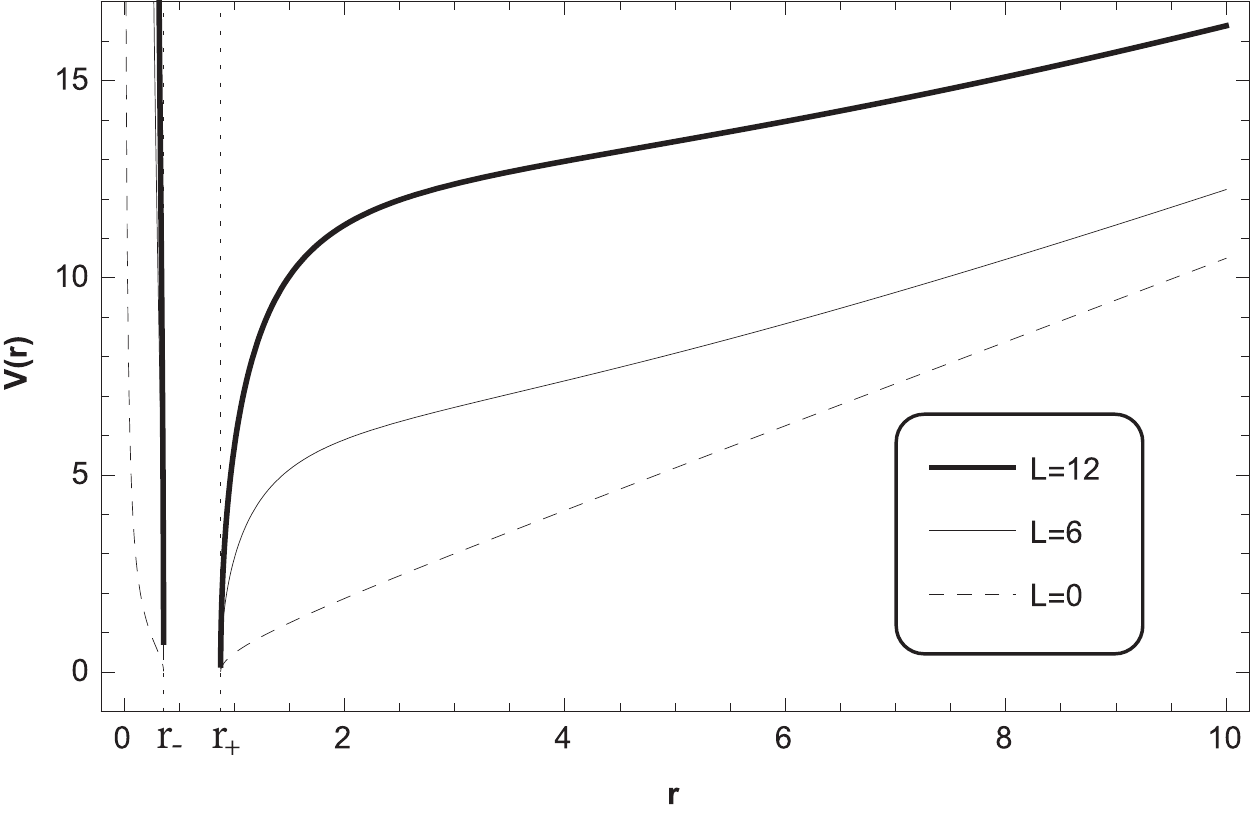}
		\includegraphics[width=60mm]{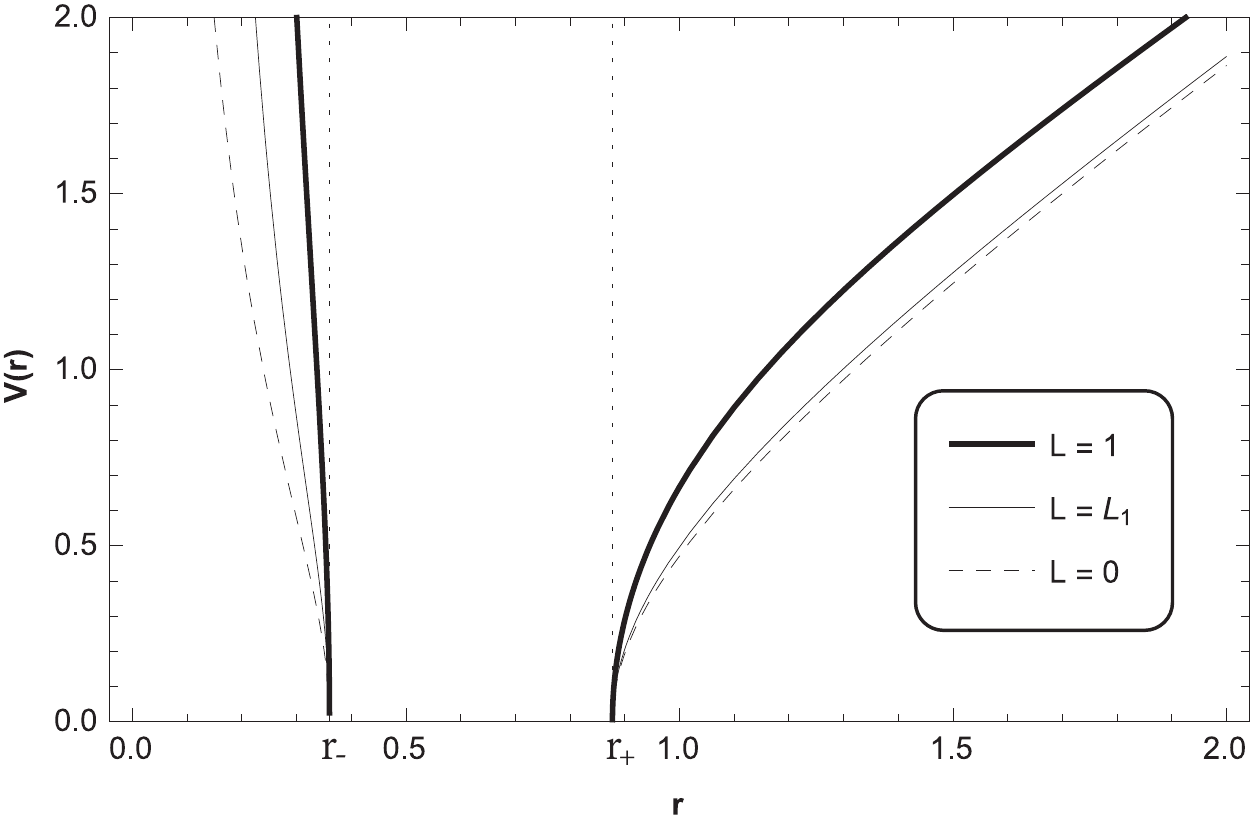}
	\end{center}
	\caption{The behavior of $V(r)$  as a function of $r$, for different values of the angular momentum of the particle $L$,
	with $M=a=b=\lambda=1$, $J=0$, $\xi=0.9$
	and $\Lambda=-1$. Here $r_+=0.878$, $r_-=0.360$, and $L_1=0.333$. } 
	\label{fpot}
\end{figure}

\begin{figure}[!h]
	\begin{center}
		\includegraphics[width=60mm]{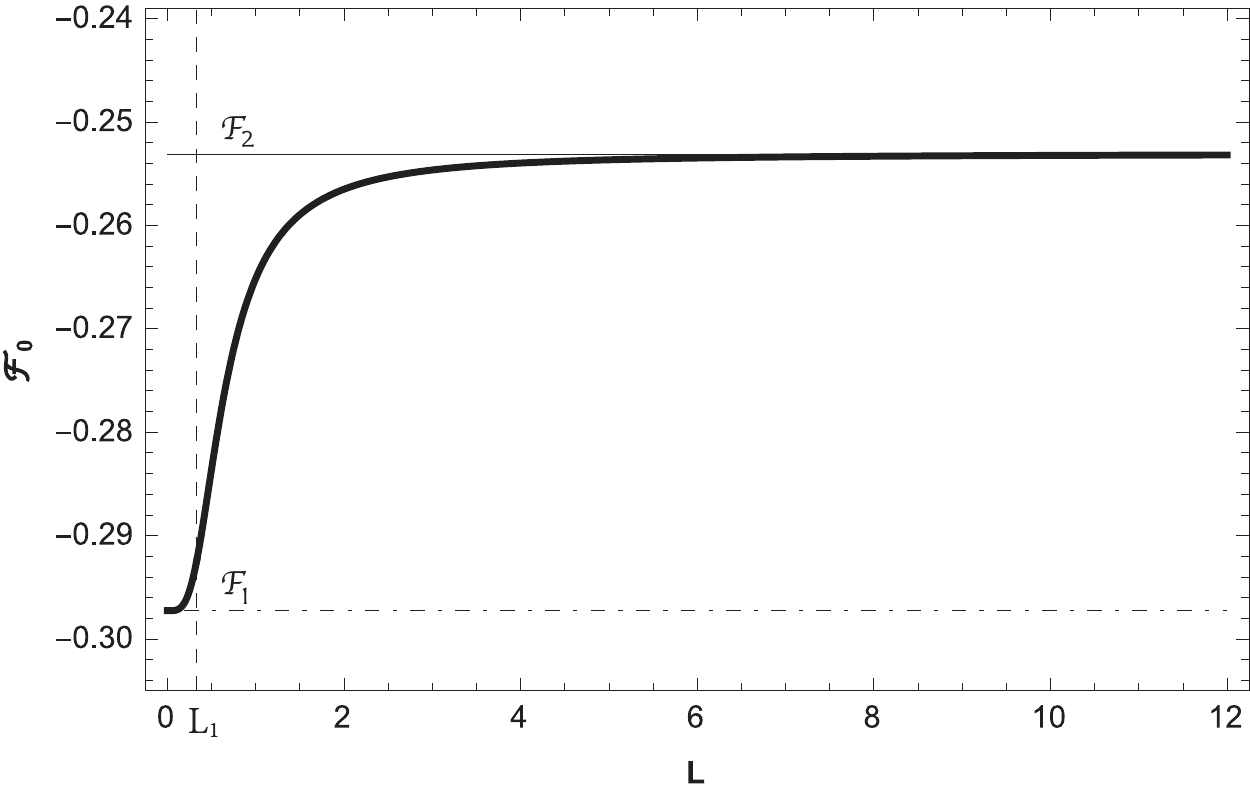}
	\end{center}
	\caption{The behavior of $\mathcal{F}_0(L)$  as a function of $L$,
	with $M=a=b=\lambda=1$, $J=0$, $\xi=0.9$, $L_1=0.333$,
	and $\Lambda=-1$. The metric function at $r=r_{ext}$ tends to $\mathcal{F}_2=-0.253$ when $L\rightarrow \infty$ and it tends to $\mathcal{F}_1=-0.297$ when $L\rightarrow 0$. Here, the function $\mathcal{F}_0(L)=\mathcal{F}[r_{ext}(L)]$, where
$\mathcal{F}_1=\mathcal{F}_0(L\rightarrow 0)=-M+\bar{J}\sqrt{(-\bar{\Lambda})}$, and
$\mathcal{F}_2=\mathcal{F}_0(L\rightarrow \infty)=-{M\over 2}+{\bar{J}^2(-\bar{\Lambda})\over 2M}$.}
	\label{flapse}
\end{figure}

In the case of non-rotating black holes, the polynomial (\ref{P12}) has analytical solution and it at $r=r_{ext}$ can be written as  
\begin{equation}
r_{ext}^6+{1\over \bar{\Lambda}}
\left({\bar{J}^2\over 4}-ML^2\right)r_{ext}^2+{\bar{J}^2\,L^2\over 2\bar{\Lambda}}=0\,,
\label{C.221}\end{equation}
where the quadratic term is null for $L=L_1={\bar{J}\over 2\sqrt{M}}$\,. The root of this polynomial is
\begin{equation}
r_{ext}=\left(\sqrt{\chi_2\over 3}\cosh\left[ {1\over 3}\cosh^{-1}\left(3\chi_3\sqrt{3\over \chi_2^3}\right) \right]\right)^{1/2}\,,\label{ext1}\\
\end{equation}
for $0<L<L_1$.
\begin{equation}
r_{ext}=\left({ \bar{J}^2\,L_1^2\over -2 \bar{\Lambda}}\right)^{1/6}=\left({ \bar{J}^4\over -8M \bar{\Lambda}}\right)^{1/6}\,, \label{ext2}\\    
\end{equation}
for $L=L_1$, and
\begin{equation}
r_{ext}=\left(\sqrt{-\chi_2\over 3}\sinh\left[ {1\over 3}\sinh^{-1}\left(3\chi_3\sqrt{3\over -\chi_2^3}\right) \right]\right)^{1/2}\,, 
\label{ext3}
\end{equation}
for  $L>L_1$, where
\begin{equation}
\chi_2=-{4\over \bar{\Lambda}}
\left({\bar{J}^2\over 4}-ML^2\right)\,,\,\, \chi_3=-{2\bar{J}^2\,L^2\over \bar{\Lambda}}\,. 
\end{equation}
In Fig. \ref{fpot} we show the behaviour of the effective potential for non-rotating black holes for large and small values of $L$,
and in Fig. \ref{flapse}, we show that the lapse function at $r=r_{ext}$ is negative for different values of $L$. Therefore, for $J=0$ we have shown that there are no circular orbits  in  the  domain $r > r_+$.

\newpage
 
\begin{figure}[!h]
	\begin{center}
		\includegraphics[width=60mm]{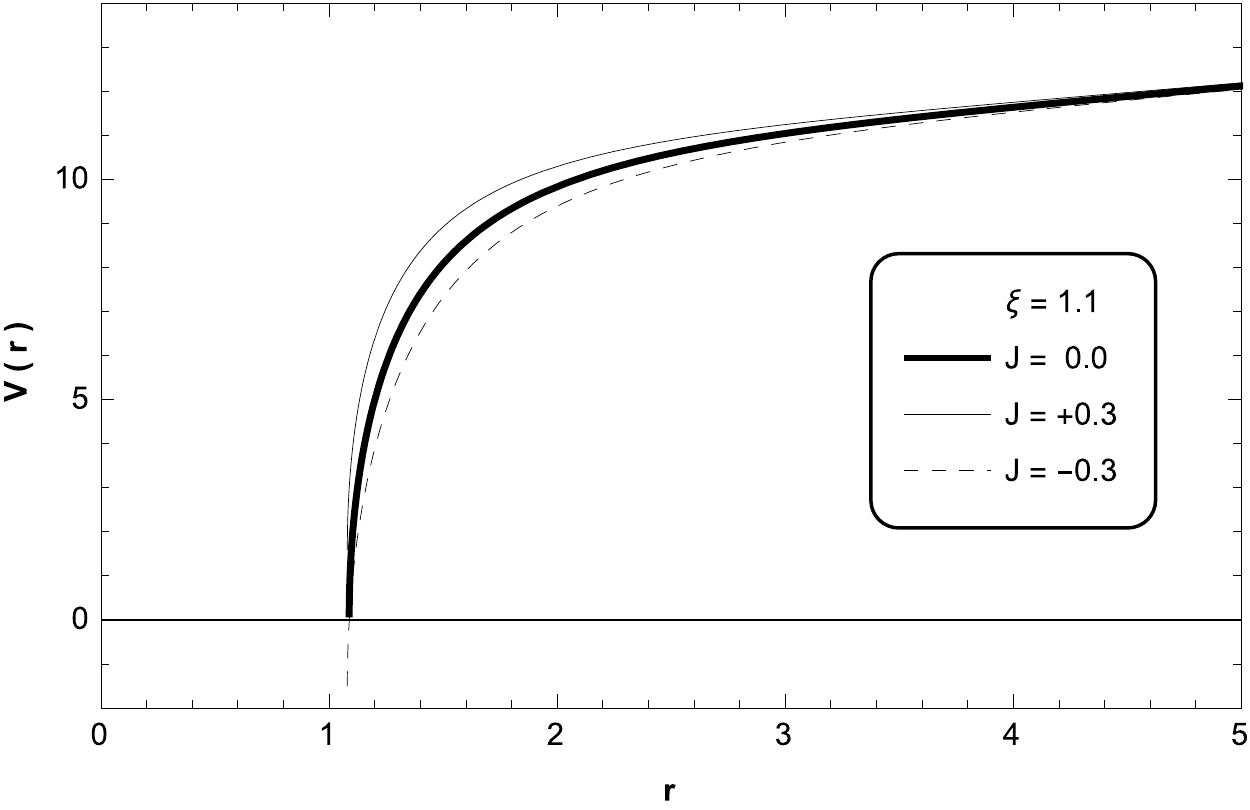}
		\includegraphics[width=60mm]{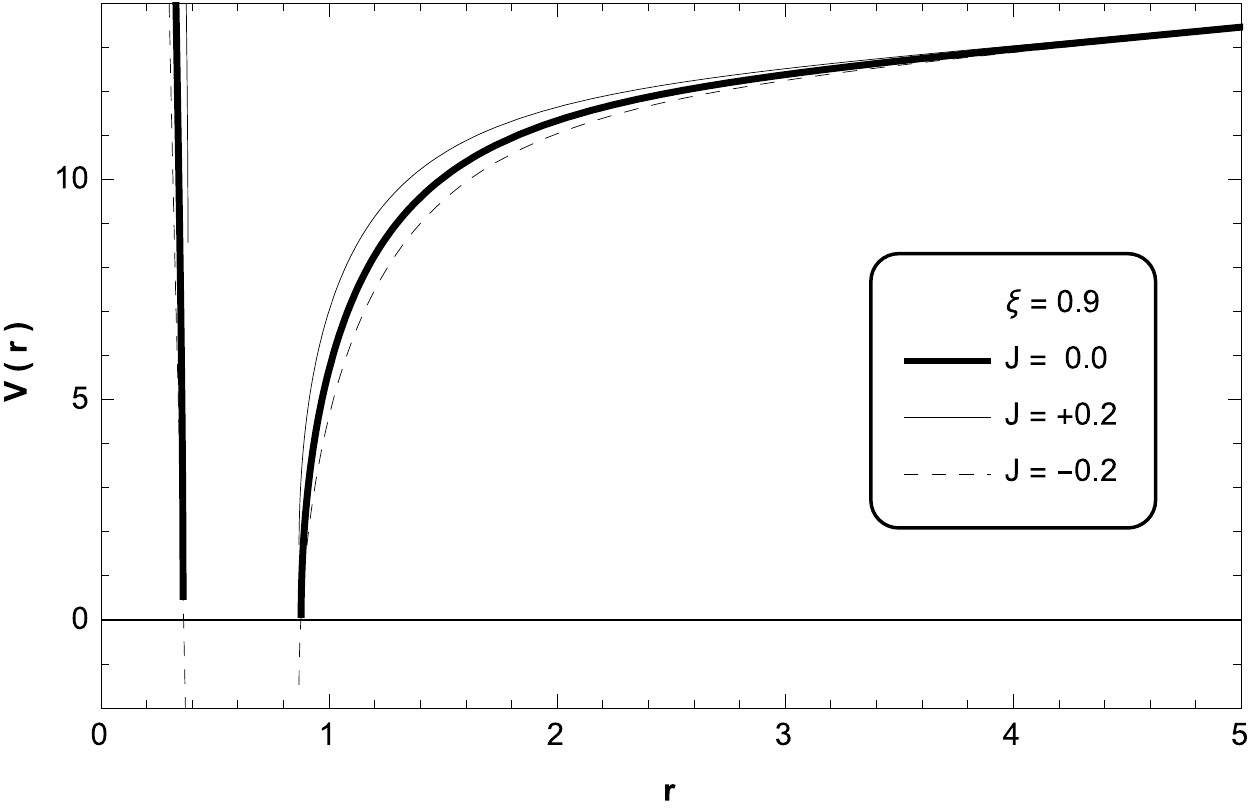}
	\end{center}
	\caption{The behaviour of the effective potential $V(r)$ as a function of $r$, for different values of the angular momentum of the black hole $J$,
	with $M=a=b=\lambda=1$, 
	$\Lambda=-1$, and $L=12$. Left panel for $\xi=1.1$, and right panel for $\xi=0.9$.}
	\label{potPP}
\end{figure}

On the other hand, in Fig. \ref{potPP}, we show that there is not circular orbit in the relevant domain $r>r_+$, for black holes with one horizon (left panel) and for black holes with an inner and outer horizon (right panel). Moreover, in this domain, the behaviour for a non-rotating black hole is similar to the behaviour for small values of $|J|$.

\subsection{Critical trajectories}
\begin{figure}[!h]
    \begin{center}
        \includegraphics[width=60mm]{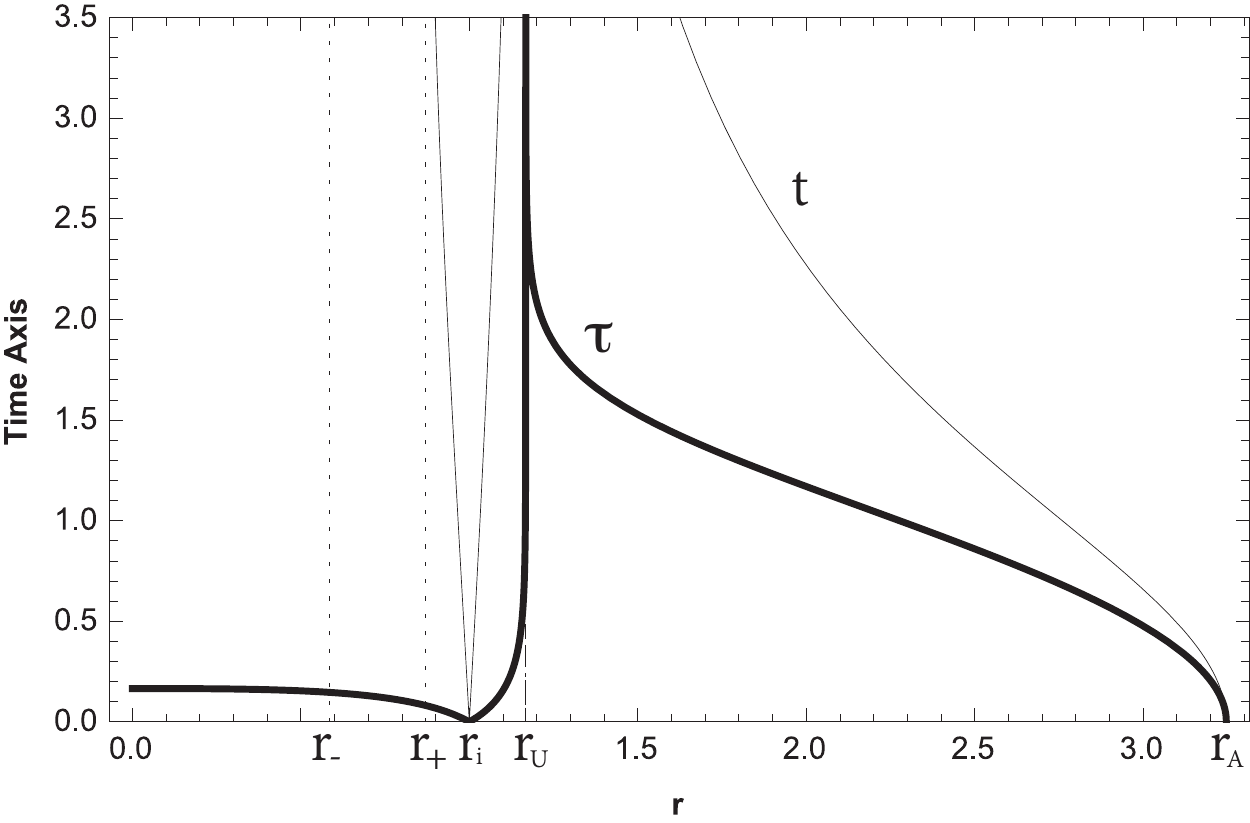}
    \end{center}
    \caption{The behaviour of the proper ($\tau$) (thick line) and coordinate ($t$) (thin line) time as a function of $r$ for critical trajectories of first and second kind with $M=a=b=\lambda=1$, $\Lambda=-1$, $J=1.2$, $\xi=1.1$, $r_-\approx 0.59$, $r_+\approx 0.87$, $r_i=1.00$, $r_U\approx 1.17$,  $r_A\approx 3.25$, and $E_U\approx 11.92$.}
    \label{crt1y2time}
\end{figure}

There are two critical orbits that approach to the unstable circular orbit asymptotically. In the first kind, the particle arises from $r=r_A$, and in the second kind, the particle starts from a finite distance $r=r_i$ bigger than the horizon radius, but smaller than the unstable radius, see Fig. \ref{crt1y2time}. By considering $\phi=t=\tau=0$, and Eq. (\ref{w.13}), the proper time for the critical orbit of first kind is
\begin{equation}
\tau(r)=-{1\over (-\bar{\Lambda})^{1/2}}\int_{r_A}^{r} {r^2\,d\,r\over (r^2-r_U^2)\sqrt{r^2_A-r^2}}~,
\end{equation}
whose solution is given by
\begin{equation}
\tau(r)={1\over \sqrt{(-\bar{\Lambda})}} \left[\psi(r)+{r_U\over \sqrt{r^2_A-r^2_U}}
\psi_U (r) \right]~,
\label{mr.3}
\end{equation}
where
\begin{equation}
\psi[r]=\tan ^{-1}\sqrt{\frac{r_A^2}{r^2}-1}\,,
\end{equation}
and
\begin{equation}
\psi_U[r]=\tanh ^{-1}\left[\frac{r_U}{r}\sqrt{\frac{r_A^2-r^2}{r_A^2-r^2_U}}\right]\,.
\end{equation}
Now, for the critical trajectories of second kind the proper time is
\begin{equation}
\tau(r)={1\over \sqrt{(-\bar{\Lambda})}} \left[\psi(r)-\psi(r_i)+{r_U\left[
    \psi_U (r)-\psi_U (r_i)\right]\over \sqrt{r^2_A-r^2_U}} \right]~,
\label{mr.3}
\end{equation}

On the other hand, by considering $\phi=t=\tau=0$, Eqs. (\ref{w.12}) and (\ref{w.13}) the coordinate time for the critical trajectory of first kind is
\begin{equation}
t(r)=-\int_{r_A}^{r}  {r^2\left[E r^2-JL/2\right]d\,r\over (-\bar{\Lambda})^{3/2}(r^2-r_+^2)(r^2-r_-^2)(r^2-r_U^2)\sqrt{r^2_A-r^2}}~,
\end{equation}
whose solution is
\begin{equation}
t(r)= {1\over (-\bar{\Lambda})^{3/2}} \sum _{j=1}^3 \tilde{k}_j \, \psi_j(r)\,,
\end{equation}
where
\begin{eqnarray}
\psi_1(r)={r_U\over \sqrt{r^2_A-r^2_U}}
\tanh ^{-1}\left[\frac{r_U}{r}\sqrt{\frac{r_A^2-r^2}{r_A^2-r^2_U}}\right]\,,\\
\psi_2(r)={r_+\over \sqrt{r^2_A-r^2_+}}
\tanh ^{-1}\left[\frac{r_+}{r}\sqrt{\frac{r_A^2-r^2}{r_A^2-r^2_+}}\right]\,,\\
\psi_3(r)={r_-\over \sqrt{r^2_A-r^2_-}}
\tanh ^{-1}\left[\frac{r_-}{r}\sqrt{\frac{r_A^2-r^2}{r_A^2-r^2_-}}\right]\,
\end{eqnarray}
and
\begin{equation}
    \tilde{k}_1= {E r_U^2-JL/2\over (r_U^2-r_+^2)(r_U^2-r_-^2)}\,, \quad
    \tilde{k}_2=-{E r_+^2-JL/2\over (r_U^2-r_+^2)(r_+^2-r_-^2)}\,, \quad
    \tilde{k}_3={E r_-^2-JL/2\over (r_U^2-r_-^2)(r_+^2-r_-^2)}\,.
\end{equation}
For the critical trajectories of second kind the coordinate time is
\begin{equation}
t(r)= {1\over (-\bar{\Lambda})^{3/2}} \sum _{j=1}^3 \tilde{k}_j \, \left[ \psi_j(r)-\psi_j(r_i)\right] \,.
\label{tc2}
\end{equation}

\begin{figure}[!h]
    \begin{center}
        \includegraphics[width=70mm]{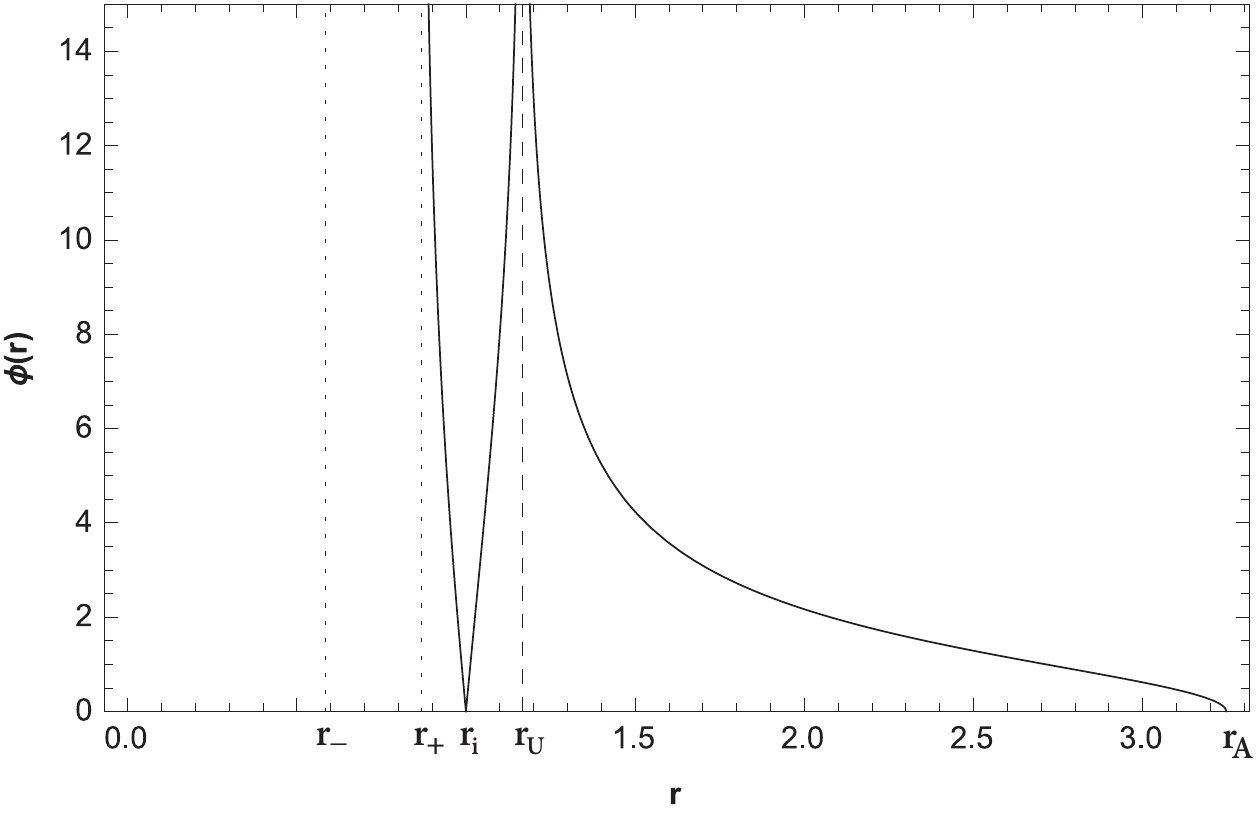}
            \includegraphics[width=60mm]{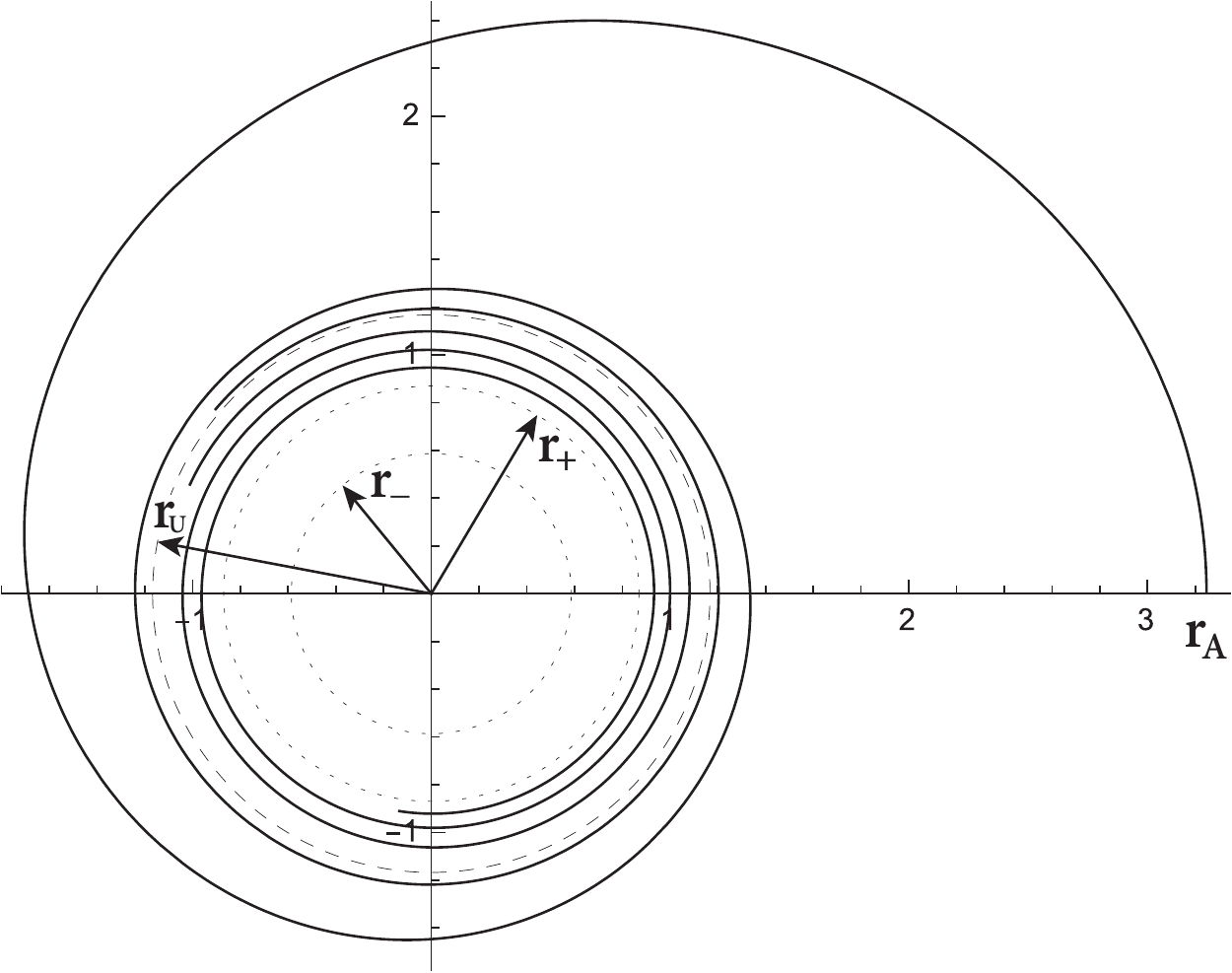}
    \end{center}
    \caption{Critical trajectory of first and second kind, for particles with  $L=12$, $M=a=b=\lambda=1$, $\Lambda=-1$, $J=1.2$, $\xi=1.1$, $E_U=11.92$, $r_-\approx 0.59$, $r_+\approx 0.87$, and $r_i=1.0$. For orbit of the first kind (thin line) the test particle arrived from $r_A= 3.25$, where $r_U\approx 1.17$  corresponds to the radius of the unstable circular orbit (dashed circle).}
    \label{fC}
\end{figure}

In Fig. \ref{crt1y2time}, we plot the behaviour of the proper and coordinate time as a function of $r$. We observe that for both times, the particle take an infinity time in to arrive to the unstable circular orbit.
Finally, by using Eq. (\ref{w.14}), and (\ref{w.13}), the angular coordinate $\phi$ for the trajectories of first kind is
\begin{equation}
\phi_C(r)= {1\over (-\bar{\Lambda})^{3/2}} \sum _{i=1}^3 \eta_i \, \psi_i(r)\,,
\end{equation}
where
\begin{equation}
\eta_1={1\over (r_U^2-r_+^2)(r_U^2-r_-^2)} \left(\frac{E_u J}{2}-LM+(-\bar{\Lambda})L\,r^2_U-\frac{L(J^2-\bar{J}^2)}{4 r_U^2}\right)\,,
\end{equation}
\begin{equation}
\eta_2={-1\over (r_U^2-r_+^2)(r_+^2-r_-^2)} \left(\frac{E_u J}{2}-LM+(-\bar{\Lambda})L\,r^2_+-\frac{L(J^2-\bar{J}^2)}{4 r_+^2}\right)\,,
\end{equation}
\begin{equation}
\eta_3={1\over (r_U^2-r_-^2)(r_+^2-r_-^2)} \left(\frac{E_u J}{2}-LM+(-\bar{\Lambda})L\,r^2_--\frac{L(J^2-\bar{J}^2)}{4 r_-^2}\right)\,.
\end{equation}
While that for critical trajectories of second kind is
\begin{equation}
\phi_C(r)= {1\over (-\bar{\Lambda})^{3/2}} \sum _{j=1}^3 \eta_j \, \left[ \psi_j(r)-\psi_j(r_i)\right] \,.
\label{tc2}
\end{equation}
In Fig. \ref{fC}, we show the behavior of the angular coordinate as a function of $r$. We can observe that for both trajectories, first and second kind, the angular coordinate diverges at the unstable circular radius.

\clearpage

%%%%%%%%%%%%%%%%%%%%%%%%%%%%%%%%%

\subsection{Motion with $L=0$}

%{\bf{For}} the {\bf{motion with $L=0$,}} 
%corresponds to a trajectory with null angular momentum $L=0$, and the particles, 
 In this case, the particles are destined to fall towards the event horizon, see Fig. \ref{f4.1}. The effective potential Eq. (\ref{tl8}) is
%we can see that for radial  particles
$V(r)=\left( -M+\frac{\bar{J}^{2}}{4\, r^2}-\bar{\Lambda}r^2\right)^{1/2}$, and  Eqs. (\ref{w.14}), (\ref{w.12}) and (\ref{w.13}) yield
\begin{eqnarray}
\label{w.13r}
&&\dot{\phi}= -{E \,J\over 2\bar{\Lambda}(r^2-r_+^2)(r^2-r_-^2)}~,\\
\label{w.12r}
&&\dot{t}= -{E \,r^2\over \bar{\Lambda}(r^2-r_+^2)(r^2-r_-^2)}~,\\
\label{w.14r2}
&&\pm \dot{r}=\sqrt{E^2+M-{\bar{J}^2\over 2 r^2}+\bar{\Lambda} \,r^2}~,
\end{eqnarray}
where the ($-$) sign for $ \dot{r}$, corresponds to particles falling into the event horizon, and the ($+$) sign corresponds to  particles that have a return point $r_0>r_+$, for $E>E_+$, given by
    \begin{equation}
    \label{r0}
    r_{0}=\sqrt{\frac{M+E^2}{-\bar{\Lambda}}} \sin \left[\frac{1}{2} \sin ^{-1}\left(\frac{\bar{J} \sqrt{-\bar{\Lambda} }}{M+E^2}\right)+\frac{\pi}{2}\right]~,
    \end{equation}
and a return point at $d_0<r_-$, given by
 \begin{equation}
    \label{r0}
    d_{0}=\sqrt{\frac{M+E^2}{-\bar{\Lambda}}} \sin \left[\frac{1}{2} \sin ^{-1}\left(\frac{\bar{J} \sqrt{-\bar{\Lambda} }}{M+E^2}\right)\right]~.
    \end{equation}

Now, choosing as initial conditions that the  particle starts at $r=r_0$, and $\phi=t=\tau=0$, the solution of  Eq. (\ref{w.14r2}) is
    \begin{equation}
    \tau(r)=\frac{1}{2 \sqrt{-\bar{\Lambda} }}\left(\sin ^{-1}\left[ \frac{M+2 \bar{\Lambda}  r^2+E^2}{\sqrt{\bar{J}^2 \bar{\Lambda} +\left(M+E^2\right)^2}}\right] -\sin ^{-1}\left[ \frac{M+2 \bar{\Lambda} r_0^2+E^2}{\sqrt{\bar{J}^2 \bar{\Lambda} +\left(M+E^2\right)^2}}\right]  \right) ~.
    \label{mr.3}
    \end{equation}
    Also, a straightforward integration of Eqs. (\ref{w.12r}) and (\ref{w.14r2}) leads to
    \begin{equation}
    t(r)={E\over 2(-\bar{\Lambda})^{3/2}}\left[\frac{r_{+}^2}{\sqrt{p(r_+)}}
    \ln \left|\frac{r_0^2-r_+^2}{r^2-r_{+}^2}\cdot\frac{F_{+}(r)}{F_{+}(r_0)}\right|-
    \frac{r_{-}^2}{\sqrt{p(r_-)}}
    \ln \left|\frac{r_0^2-r_-^2}{r^2-r_{-}^2}\cdot\frac{F_{-}(r)}{F_{-}(r_0)}\right|
    \right]~,
    \label{mr.5}
    \end{equation}
    where
    \begin{eqnarray}
    \label{Fr}
    &&F_{\pm}(r)=2p(r_{\pm})+\left( {E^2+M \over -\bar{\Lambda}}-2\,r_{\pm}^2\right) (r^2-r_{\pm}^2)+2\sqrt{p(r_{\pm})}\sqrt{P_{\pm}(r)}~,\\
    \label{pr}
    &&p(r)=-r^4+{E^2+M \over -\bar{\Lambda}}r^2-{\bar{J}^2\over 4(-\bar{\Lambda})} ~,\\
        \label{Pr}
    &&P_{\pm}(r)=p(r_{\pm})+\left( {E^2+M \over -\bar{\Lambda}}-2\,r_{\pm}^2\right) (r^2-r_{\pm}^2)-(r^2-r_{\pm}^2)^2 ~.
    \end{eqnarray}

\begin{figure}[!h]
        \begin{center}
            \includegraphics[width=60mm]{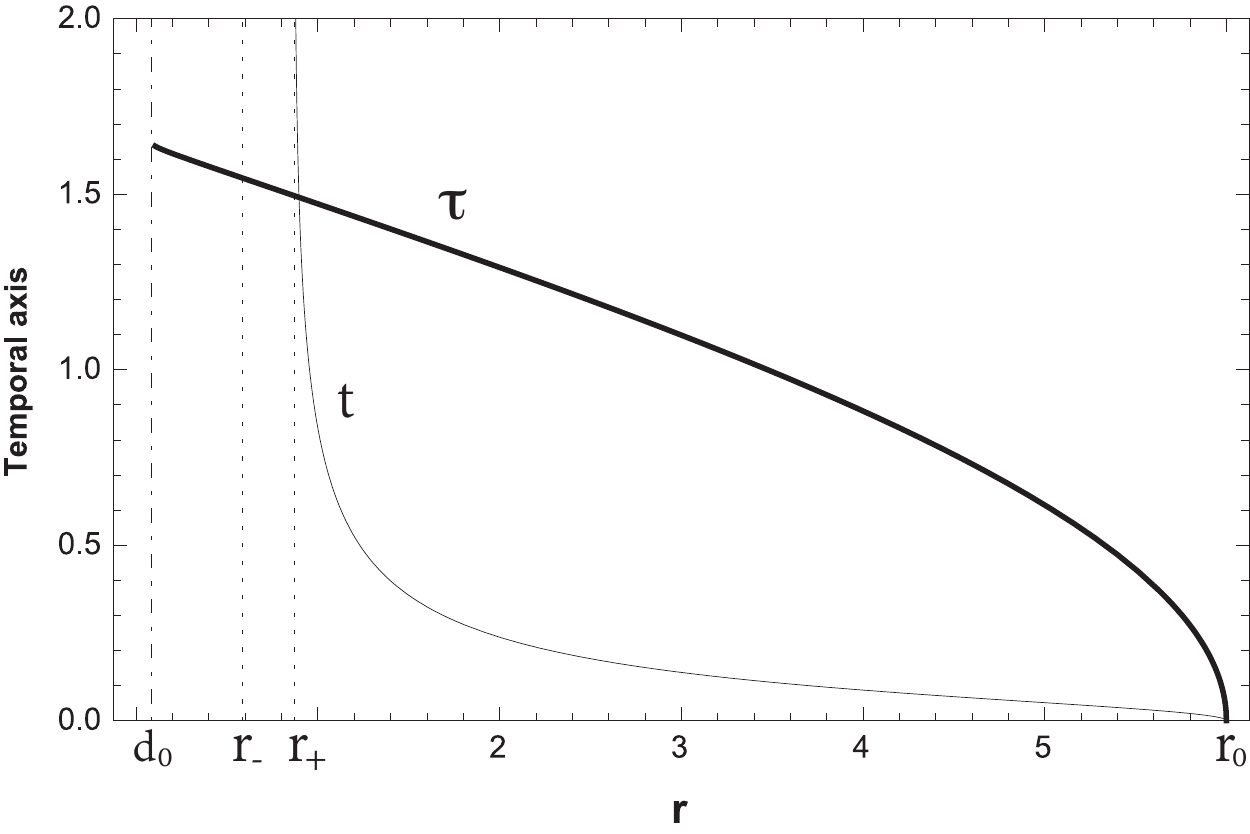}
        \end{center}
        \caption{The behaviour of the coordinate time $(t)$ and the proper time   $(\tau)$ along an unbounded time-like radial geodesic described by a test particle, starting at $r_0=6$ and falling towards the singularity,
        %or going towards infinity,
        for $L=0$, $M=a=b=\lambda=1$, $\Lambda=-1$, $E=5.63$, $J=1.2$, $\xi=1.1$, $r_+\approx 0.87$, $d_0=0.085$, and $r_{-} \approx 0.59$.}
        \label{tauytr}
    \end{figure}

\begin{figure}[!h]
        \begin{center}
            \includegraphics[width=60mm]{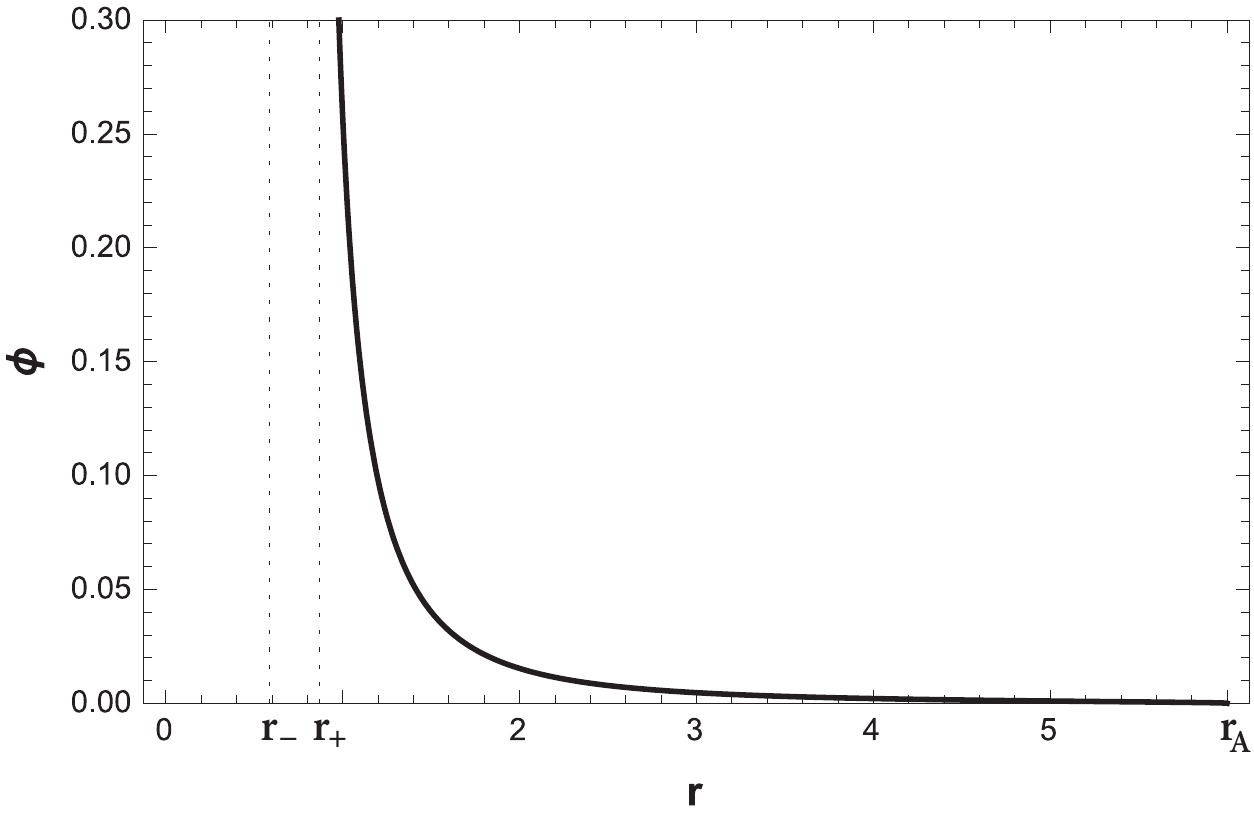}\hspace{5mm}
            \includegraphics[width=80mm]{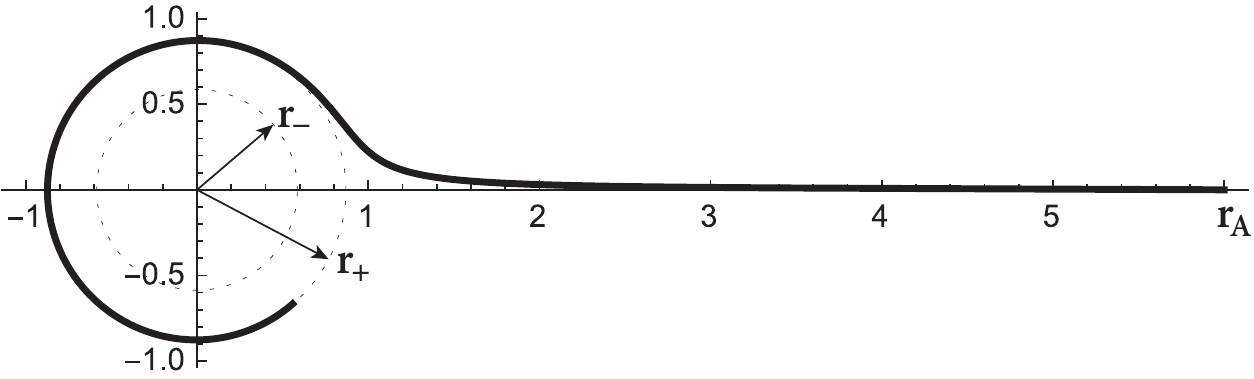}\hspace{5mm}
        \end{center}
        \caption{The behavior of the coordinate $\phi(r)$, starting at $r_0=6$ and falling towards the singularity with  $L=0$, $M=a=b=\lambda=1$, $\Lambda=-1$, $J=1.2$, $\xi=1.1$, $r_{-} \approx 0.59$ and $r_+=0.87$.
            In the polar plot, the trajectory approaching the horizon will spiral around the black hole an infinite number of times.}
        \label{phiradr}
    \end{figure}

In Fig. \ref{tauytr}, we show the behaviour of the proper and coordinate time as a function of $r$, we can observe that the particle arrives to the event horizon in a finite proper time, then the particle does not reach the singularity, due to the existence of a return point $d_0$ inside $r_-$, see Fig. \ref{f4.11}. Concerning to the coordinate time, the particle arrives in an  infinite coordinate time to the singularity.
Finally, the solution for the angular coordinate  $\phi$ is
        \begin{equation}
    \phi(r)={E\,J\over 4(-\bar{\Lambda})^{3/2}}\left[\frac{1}{\sqrt{p(r_+)}}
    \ln \left|\frac{r_0^2-r_+^2}{r^2-r_{+}^2}\cdot\frac{F_{+}(r)}{F_{+}(r_0)}\right|-
    \frac{1}{\sqrt{p(r_-)}}
    \ln \left|\frac{r_0^2-r_-^2}{r^2-r_{-}^2}\cdot\frac{F_{-}(r)}{F_{-}(r_0)}\right|
    \right]~,
    \label{mr.5b}
    \end{equation}
  where we have used Eq. (\ref{w.13r}) and Eq. (\ref{w.14r2}). In Fig. \ref{phiradr}, we plot the behaviour of the angular coordinate, where we observe that the angular coordinate becomes infinity at the event horizon.

\newpage

\section{Remarks and conclusions}
\label{conclusion}

In this work, we studied the motion of particles in the background of a rotating three-dimensional Ho\v{r}ava AdS black hole described by a Lorentz-violating version of the BTZ black hole,
and we calculated the time like geodesics, which posses a rich structure and allow different kinds of trajectories for the particles. This work along with the null geodesic described in Ref. \cite{Gonzalez:2019xfr}, complement the geodesic structure for the rotating three-dimensional Ho\v{r}ava AdS black hole.  For direct orbits,  we have shown the existence of planetary orbits, where we have obtained an exact solution and we have determined the periods of revolution. Also, for circular orbits, we have shown the existence of the stable and unstable circular orbits, and we determined the periods of revolution, as well as, the epicycle frequency for the stable circular orbit. Also, critical orbits of first and second kind that approach to the unstable circular orbit asymptotically. For the motion with $0\leq L <L_C$, the 
trajectories 
are all bounded, and for $L=0$, we have obtained exact solutions. However, their counterpart, i.e, the BTZ metric
 allows geodesics for massive particles that always fall into the event horizon and no stable orbits are possible;
thereby, the differences observed with respect to the BTZ metric could be attributed to the breaking of the Lorentz invariance. 
On the other hand, for retrograde orbits we have shown the existence of second kind orbits, similar to the behaviour observed for retrograde orbits in a BTZ black hole background \cite{Cruz:1994ir}. In addition, by comparing both orbits, direct and retrograde, in a rotating three-dimensional Ho\v{r}ava AdS black hole it is possible to observe trajectories with $E\leq 0$ for retrograde orbits which is not possible for direct orbits.

Therefore, the Lorentz-violating version of the BTZ black hole,
turn on an effective potential with a more rich structure allowing different kind of orbits. As it was shown \cite{Gonzalez:2019xfr}, for photons new kinds of orbits are allowed, such as unstable circular orbits and trajectories of first kind. While that for particles the planetary and circular orbits are allowed, which does not occurs in the BTZ background. In this way, the breaking of the Lorentz invariance could generate orbits that could not occur in invariant Lorentz theories.

\section*{Acknowledgments}

We thank the referee for his/her careful review of the manuscript and his/her valuable comments and suggestions which helped us to improve the manuscript. Y.V. acknowledge support by the Direcci\'on de Investigaci\'on y Desarrollo de la Universidad de La Serena, Grant No. PR18142.


\begin{thebibliography}{99}

%\cite{Banados:1992wn}

\bibitem{Banados:1992wn}  M.~Banados, C.~Teitelboim and J.~Zanelli,
``The Black hole in three-dimensional space-time,''
Phys.\ Rev.\ Lett.\ \textbf{69} (1992) 1849  [hep-th/9204099].
%%CITATION = HEP-TH/9204099;%%
%1668 citations counted in INSPIRE as of 20 Apr 2014

\bibitem{Carlip:1995qv}
  S.~Carlip,
  ``The (2+1)-Dimensional black hole,''
  Class.\ Quant.\ Grav.\  {\bf 12}, 2853 (1995)
 %  doi:10.1088/0264-9381/12/12/005
  [gr-qc/9506079].

%\cite{Farina:1993xw}
\bibitem{Farina:1993xw}
  C.~Farina, J.~Gamboa and A.~J.~Segui-Santonja,
  ``Motion and trajectories of particles around three-dimensional black holes,''
  Class.\ Quant.\ Grav.\  {\bf 10}, L193 (1993)
%  doi:10.1088/0264-9381/10/11/001
  [gr-qc/9303005].
  %%CITATION = doi:10.1088/0264-9381/10/11/001;%%
  %21 citations counted in INSPIRE as of 24 Sep 2019



%\cite{Cruz:1994ir}
\bibitem{Cruz:1994ir}
  N.~Cruz, C.~Martinez and L.~Pena,
  ``Geodesic structure of the (2+1) black hole,''
  Class.\ Quant.\ Grav.\  {\bf 11}, 2731 (1994)
%  doi:10.1088/0264-9381/11/11/014
  [gr-qc/9401025].
  %%CITATION = doi:10.1088/0264-9381/11/11/014;%%
  %52 citations counted in INSPIRE as of 14 Jul 2019

    %\cite{Sotiriou:2011dr}
\bibitem{Sotiriou:2011dr}
  T.~P.~Sotiriou, M.~Visser and S.~Weinfurtner,
  ``Lower-dimensional Horava-Lifshitz gravity,''
  Phys.\ Rev.\ D {\bf 83}, 124021 (2011)
  %doi:10.1103/PhysRevD.83.124021
  [arXiv:1103.3013 [hep-th]].
  %%CITATION = doi:10.1103/PhysRevD.83.124021;%%
  %29 citations counted in INSPIRE as of 30 Sep 2019

  %\cite{Sotiriou:2014gna}
\bibitem{Sotiriou:2014gna}
 T.~P.~Sotiriou, I.~Vega and D.~Vernieri,
 ``Rotating black holes in three-dimensional Ho\v{r}ava gravity,''
  Phys.\ Rev.\ D {\bf 90}, no. 4, 044046 (2014)
 % doi:10.1103/PhysRevD.90.044046
  [arXiv:1405.3715 [gr-qc]].
  %%CITATION = doi:10.1103/PhysRevD.90.044046;%%
  %41 citations counted in INSPIRE as of 01 Jul 2019




%\cite{Horava:2009uw}
\bibitem{Horava:2009uw}
  P.~Horava,
  ``Quantum Gravity at a Lifshitz Point,''
  Phys.\ Rev.\ D {\bf 79}, 084008 (2009)
  %doi:10.1103/PhysRevD.79.084008
  [arXiv:0901.3775 [hep-th]].
  %%CITATION = doi:10.1103/PhysRevD.79.084008;%%
  %1788 citations counted in INSPIRE as of 30 Sep 2019






%\cite{Becar:2019hwk}
\bibitem{Becar:2019hwk}
  R.~Bécar, P.~A.~González, E.~Papantonopoulos and Y.~Vásquez,
  ``Quasinormal modes of three-dimensional rotating Ho$\check r$ava AdS black hole and the approach to thermal equilibrium,''
  Eur.\ Phys.\ J.\ C {\bf 80}, no. 7, 600 (2020)
  %doi:10.1140/epjc/s10052-020-8169-2
  [arXiv:1906.06654 [gr-qc]].
  %%CITATION = doi:10.1140/epjc/s10052-020-8169-2;%%
  %4 citations counted in INSPIRE as of 21 Jul 2020




 %\cite{Becar:2020hix}
\bibitem{Becar:2020hix}
  R.~Bécar, P.~A.~González and Y.~Vásquez,
  ``Collision of particles near a three-dimensional rotating Ho\v{r}ava AdS black hole,''
  arXiv:2002.04421 [gr-qc].
  %%CITATION = ARXIV:2002.04421;%%


%\cite{Gonzalez:2019xfr}
\bibitem{Gonzalez:2019xfr}
  P.~A.~González, M.~Olivares, E.~Papantonopoulos and Y.~Vásquez,
  ``Motion and trajectories of photons in a three-dimensional rotating Hořava-AdS black hole,''
  Phys.\ Rev.\ D {\bf 101}, no. 4, 044018 (2020)
 % doi:10.1103/PhysRevD.101.044018
  [arXiv:1912.00946 [gr-qc]].
  %%CITATION = doi:10.1103/PhysRevD.101.044018;%%
  %1 citations counted in INSPIRE as of 24 Apr 2020


  %\cite{Fernando:2003gg}
\bibitem{Fernando:2003gg}
  S.~Fernando, D.~Krug and C.~Curry,
  ``Geodesic structure of static charged black hole solutions in 2+1 dimensions,''
  Gen.\ Rel.\ Grav.\  {\bf 35}, 1243 (2003).
  %doi:10.1023/A:1024449824938
  %%CITATION = doi:10.1023/A:1024449824938;%%
  %13 citations counted in INSPIRE as of 24 Jul 2020



  %\cite{Cruz:2013ufa}
\bibitem{Cruz:2013ufa}
  N.~Cruz, M.~Olivares and J.~R.~Villanueva,
  ``Geodesic Structure of Lifshitz Black Holes in 2+1 Dimensions,''
  Eur.\ Phys.\ J.\ C {\bf 73}, 2485 (2013)
 % doi:10.1140/epjc/s10052-013-2485-8
  [arXiv:1305.2133 [gr-qc]].
  %%CITATION = doi:10.1140/epjc/s10052-013-2485-8;%%
  %11 citations counted in INSPIRE as of 24 Jul 2020

  %\cite{Kazempour:2017gho}
\bibitem{Kazempour:2017gho}
  S.~Kazempour and S.~Soroushfar,
  ``Investigation the geodesic motion of three dimensional rotating black holes,''
  Chin.\ J.\ Phys.\  {\bf 65}, 579 (2020)
%  doi:10.1016/j.cjph.2020.04.004
  [arXiv:1709.06541 [gr-qc]].
  %%CITATION = doi:10.1016/j.cjph.2020.04.004;%%

  %\cite{Panotopoulos:2020zvi}
\bibitem{Panotopoulos:2020zvi}
  G.~Panotopoulos,
  ``Orbits of test particles in three-dimensional Maxwell–Dilaton spacetime: exact analytical solution to the geodesic equation,''
  Gen.\ Rel.\ Grav.\  {\bf 52}, no. 6, 54 (2020).
 % doi:10.1007/s10714-020-02706-x
  %%CITATION = doi:10.1007/s10714-020-02706-x;%%






  %\cite{Jacobson:2010mx}
\bibitem{Jacobson:2010mx}
  T.~Jacobson,
  ``Extended Horava gravity and Einstein-aether theory,''
  Phys.\ Rev.\ D {\bf 81}, 101502 (2010)
  Erratum: [Phys.\ Rev.\ D {\bf 82}, 129901 (2010)]
 % doi:10.1103/PhysRevD.82.129901, 10.1103/PhysRevD.81.101502
  [arXiv:1001.4823 [hep-th]].
  %%CITATION = doi:10.1103/PhysRevD.82.129901, 10.1103/PhysRevD.81.101502;%%
  %193 citations counted in INSPIRE as of 01 Jun 2020

   \bibitem{Cropp:2013sea}

  B.~Cropp, S.~Liberati, A.~Mohd and M.~Visser,
  ``Ray tracing Einstein-Æther black holes: Universal versus Killing horizons,''
  Phys.\ Rev.\ D {\bf 89}, no. 6, 064061 (2014)
 % doi:10.1103/PhysRevD.89.064061
  [arXiv:1312.0405 [gr-qc]].
  %%CITATION = doi:10.1103/PhysRevD.89.064061;%%

   \bibitem{Zhu:2019ura}
  T.~Zhu, Q.~Wu, M.~Jamil and K.~Jusufi,
  ``Shadows and deflection angle of charged and slowly rotating black holes in Einstein-Æther theory,''
  Phys.\ Rev.\ D {\bf 100}, no. 4, 044055 (2019)
  %doi:10.1103/PhysRevD.100.044055
  %[arXiv:1906.05673 [gr-qc]].
  %%CITATION = doi:10.1103/PhysRevD.100.044055;%%
  %\cite{Becar:2019hwk}


  \bibitem{chandra}
Chandrasekhar S.:
The Mathematical Theory of Black Holes.
Oxford University Press, New York (1983).



  %\cite{Deruelle:1974zy}
\bibitem{Deruelle:1974zy}
  N.~Deruelle and R.~Ruffini,
  ``Quantum and classical relativistic energy states in stationary geometries,''
  Phys.\ Lett.\  {\bf 52B}, 437 (1974).
%  doi:10.1016/0370-2693(74)90119-1
  %%CITATION = doi:10.1016/0370-2693(74)90119-1;%%
  %64 citations counted in INSPIRE as of 30 Sep 2019






%\cite{RamosCaro:2011wx}
\bibitem{RamosCaro:2011wx}
  J.~Ramos-Caro, J.~F.~Pedraza and P.~S.~Letelier,
  ``Motion around a Monopole + Ring system: I. Stability of Equatorial Circular Orbits vs Regularity of Three-dimensional Motion,''
  Mon.\ Not.\ Roy.\ Astron.\ Soc.\  {\bf 414}, 3105 (2011)
  %doi:10.1111/j.1365-2966.2011.18618.x
  [arXiv:1103.4616 [astro-ph.EP]].
  %%CITATION = doi:10.1111/j.1365-2966.2011.18618.x;%%
  %14 citations counted in INSPIRE as of 30 May 2020


\end{thebibliography}
\end{document}